\documentclass[12pt]{article}

\pdfoutput=1

\usepackage[top=80pt,bottom=85pt,left=85pt,right=85pt]{geometry}
\usepackage{amssymb}
\usepackage{amsmath}
\usepackage{setspace}
\usepackage{sectsty}
\usepackage{cite}
\usepackage{bbm}
\usepackage[utf8]{inputenc}
\usepackage[T1]{fontenc}
\usepackage{comment}
\usepackage[english]{babel}
\usepackage{afterpage}

\usepackage[usenames]{xcolor}
\usepackage{graphicx,subcaption}
\usepackage{setspace}
\usepackage{float}
\usepackage{booktabs}
\usepackage[debug,pageanchor=false]{hyperref}
\definecolor{link}{rgb}{.8,.15,.1}
\definecolor{pigment}{rgb}{0.36, 0.54, 0.66}
\definecolor{pigment2}{rgb}{0.19, 0.55, 0.91}
\definecolor{pigment3}{rgb}{0.2, 0.2, 0.6}
\definecolor{light-gray}{gray}{0.75}
\hypersetup{colorlinks=true,linkcolor=link,citecolor=link,urlcolor=link,linktocpage}

\usepackage{array,multirow,booktabs,longtable}
\usepackage{mathrsfs}
\usepackage{tikz-cd} 
\usetikzlibrary{backgrounds, arrows,calc,shapes,decorations.pathreplacing, decorations.markings, automata,positioning}
\tikzset{%
  >={Latex[width=2mm,length=2mm]},
            base/.style = {rectangle, rounded corners, draw=black,
                           minimum width=4cm, minimum heigwht=1cm,
                           text centered, font=\sffamily},
  activityStarts/.style = {base, fill=orange!15},
       startstop/.style = {base, fill=orange!15},
    activityRuns/.style = {base, fill=orange!15},
         process/.style = {base, minimum width=2.5cm, fill=orange!15,
                           font=\ttfamily},
}
\newcommand{\red}[1]{}
\tikzset{
        cvertex/.style={circle,draw=black,inner sep=1pt,outer sep=3pt},
        vertex/.style={circle,fill=black,inner sep=1pt,outer sep=3pt},
        star/.style={circle,fill=yellow,inner sep=0.75pt,outer sep=0.75pt},
        tvertex/.style={inner sep=1pt,font=\scriptsize},
        gap/.style={inner sep=0.5pt,fill=white}}

\tikzstyle{mybox} = [draw=black, fill=blue!10, very thick,
    rectangle, rounded corners, inner sep=10pt, inner ysep=20pt]
\tikzstyle{boxtitle} =[fill=blue!50, text=white,rectangle,rounded corners]

\newcommand{\rr}{\mathbb{R}}
\newcommand{\cc}{\mathbb{C}}
\newcommand{\zz}{\mathbb{Z}}
\newcommand{\pp}{\mathbb{P}} 
\newcommand{\Ntwo}{{\mathcal{N}=2}}

\def\cC{\mathbb{C^*}}

\newcommand{\todo}[1]{}
\renewcommand{\todo}[1]{{\color{red} TODO: {#1}}}
\renewcommand{\red}[1]{{\color{red} {#1}}}

\newcommand{\be}{\begin{equation}}  
\newcommand{\ee}{\end{equation}}  
\newcommand{\bea}{\begin{align}}
\newcommand{\eea}{\end{align}}
\newcommand{\bp}{\begin{bmatrix*}[r]}  
\newcommand{\ep}{\end{bmatrix*}}  
\newcommand{\bpp}{\begin{bmatrix}}  
\newcommand{\epp}{\end{bmatrix}}  
\newcommand{\bcd}{\begin{center}
\begin{tikzcd}}
\newcommand{\ecd}{\end{tikzcd} \end{center}}

\makeatletter
\@addtoreset{equation}{section}
\makeatother
\begin{document}

% titlepage 

\begin{titlepage}

\begin{center}

\vskip .3in \noindent

{\Large \bf{The role of U(1)'s in 5d theories, Higgs branches, and geometry}}

\bigskip\bigskip

Andr\'es Collinucci$^a$ and Roberto Valandro$^b$ \\

\bigskip

%Version of \DTMNow~(w.r.t. GMT)

\bigskip
{\footnotesize
 \it

$^a$ Service de Physique Th\'eorique et Math\'ematique, Universit\'e Libre de Bruxelles and \\ International Solvay Institutes, Campus Plaine C.P.~231, B-1050 Bruxelles, Belgium\\
\vspace{.25cm}
$^b$ Dipartimento di Fisica, Universit\`a di Trieste, Strada Costiera 11, I-34151 Trieste, Italy \\%and \\
%\vspace{.25cm}
and INFN, Sezione di Trieste, Via Valerio 2, I-34127 Trieste, Italy	
}

\vskip .5cm
{\scriptsize \tt collinucci dot phys at gmail dot com \hspace{1cm}  roberto dot valandro at ts dot infn dot it}

\vskip 1cm
     	{\bf Abstract }
\vskip .1in
\end{center}
We explore the Higgs branches of five-dimensional $\mathcal{N}=1$ quiver gauge theories at finite coupling from the paradigm of M-theory on local Calabi-Yau threefolds described as $\mathbb{C}^\ast$-fibrations over local K3's. By properly counting local deformations of singularities, we find results compatible with \emph{unitary} as opposed to \emph{special unitary} gauge groups.
We interpret these results by dualizing to both IIA on local K3's with D6-branes, and to IIB with 5-branes.
Finally, we find that, by compactifying the $\mathbb{C}^\ast$-fibers to tori, a well-known St\"uckelberg mechanism eliminates Abelian factors, and provides missing Higgs branch moduli in a very interesting way. This is also explained from the dual IIA and IIB viewpoints.

\noindent

\vfill
\eject

\end{titlepage}

% end titlepage

\tableofcontents

\newpage 
%%%%%%%%%%%%%%%%%%%%%%%%%%%
\section{Introduction} % sec (intro)
\label{sec:intro}
%%%%%%%%%%%%%%%%%%%%%%%%%%%
Five-dimensional quiver gauge theories with eight supercharges ($\mathcal{N}=1, d=5$) can be engineered by putting M-theory on non-compact singular Calabi-Yau varieties. This paradigm has been used originally in \cite{Intriligator:1997pq} to build Seiberg's SCFT's \cite{Seiberg:1996bd} as their UV fixed points.

The dictionary between the data of a \emph{compact} Calabi-Yau (CY) threefold $X_3$ and a theory $T_{X_3}$ is by now well-established and detailed, has been widely used to further classify possible 5d SCFT's in numerous works such as \cite{Apruzzi:2019kgb,Apruzzi:2019vpe,Bhardwaj:2020gyu,Bhardwaj:2018vuu,Bhardwaj:2018yhy,Bhardwaj:2019xeg,Apruzzi:2019opn,Apruzzi:2019enx,Closset:2018bjz,Morrison:2020ool, Albertini:2020mdx,Jefferson:2018irk,Bhardwaj:2019jtr,Saxena:2019wuy,Jefferson:2017ahm,Eckhard:2020jyr,Closset:2019juk}, just to name a few. Two of its entries at gross level are the following: 
   \be \begin{array}{c|c}
    X_3 & T_{X_3}\\ \hline
    b_2 & {\rm dim}_{\mathbb{R}} \mathcal{C} \\
    b_3 & {\rm dim}_{\mathbb{C}} \mathcal{H}
    \end{array} 
    \ee
where $\mathcal{C}$ and $\mathcal{H}$ are the Coulomb and Higgs branches, respectively, and the $b_i$ are Betti numbers. If we want to decouple gravity, however, we should work with \emph{non-compact} threefolds, in which case we have to be more precise about what we mean by $b_2$ and $b_3$. By now, the standard practice is to replace $b_2$ by the number of compact divisors (exploiting Poincar\'e duality), and to work mainly with isolated singularities, whereby one counts the number of local deformations:
\begin{eqnarray}
b_2& \longrightarrow & \#\, \text {compact divisors}\:,\nonumber\\
b_3& \longrightarrow & \#\, \text {local deformations}\:. \nonumber
\end{eqnarray}
From this viewpoint, an SCFT can be deformed by activating appropriate K\"ahler moduli, triggering an RG flow to a weakly coupled quiver gauge theory with $SU$, $Sp$ or $SO$ gauge groups.

In this paper, we will analyze a broad class of threefolds with non-isolated singularities. Usually, these are expected to describe SCFT's which, under mass deformations, should give rise to \emph{special unitary} quivers. We will show that, under appropriate conditions, the same M-theory/IIA setups can give rise to quivers with unitary gauge groups, including fully Abelian theories, which are usually ruled out in the literature.

Usually, Abelian factors are not taken into account in 5d studies because they are expected to hit Landau poles. However, in the string theory setups we are considering, Abelian gauge groups are always accompanied by `instanton-particles' that become massless at the infinite coupling point, which might explain why they are consistently embedded. From a purely gauge theory perspective, we do not expect instantons in Abelian theories, except when an appropriate non-commutativity parameter is present. However, in string theory it is well-known that a single D6 can form a bound state at threshold with a D2-brane\footnote{In that case, the D2 is not interpreted as a vector bundle on the D6 worldvolume, but as an appropriate object of the bounded derived category of coherent sheaves.}. This will provide us with the necessary object.

Our motivation for this paper stems from the observation that, looking at M-theory singularities in the algebro-geometric framework, we  counted Higgs branch moduli at weak coupling under the assumption of special unitary gauge groups, and find exactly one missing quaternionic dimension per gauge group factor. This is consistent with the hypothesis that all gauge groups are \emph{unitary}.

The crux of the matter is that non-compact threefolds can support normalizable harmonic two-forms that are not Poincar\'e dual to compact four-cycles, provided a Taub-NUT-type metric is chosen. This is not new, as it has been known since the eighties that ALF spaces have this property \cite{Ruback:1986ag}.

To be precise, we will study local CY threefolds that are $\mathbb{C}^*$-fibrations over local K3's. These threefolds have the asset that they can easily be reduced to type IIA string theory with D6-branes wrapping holomorphic curves, where the quiver gauge theory can be read off easily: D6-branes on holomorphic compact curves give $U(N)$ gauge nodes, and D6-branes on non-compact curves give flavors. A large subclass of these examples can be torically realized, which allows us to use the methods developed in \cite{Closset:2018bjz}.

Having established these facts, we will show a mechanism to turn unitary quivers into special unitary quivers in type IIA string theory. By compactifying a transverse dimension on a circle, a St\"uckelberg mechanism  will render $U(1)$ factors massive. Correspondingly, the dimensions of the Higgs branches are expected to jump upwards. We will show that, indeed, the extra dimensions come from bulk supergravity moduli that are not accounted for by algebraic geometry alone. Without the circle compactification, these moduli are 6-dimensional, and hence considered as non-dynamical from a 5d viewpoint. A similar mechanism is considered in a different context in \cite{Hanany:1997gh, Brunner:1997gf}.

Note, that for strong string coupling, we expect $U(1)$ factors to drop out on their own, for completely different reasons, which are unclear from a string theory perspective (as remarked in \cite{Closset:2018bjz}). In M-theory, such a limit corresponds to having an ALE fibration~(as opposed to ALF), which makes us lose one normalizable two-form per gauge node.

Finally, we will give an interpretation of our results from the IIB 5-brane web perspective. In that case, we will find that if we T-dualize IIA on an ALF space at finite $g_s^{IIA}$, and take the ALE limit, we will land in IIB with $g_s^{IIB} \rightarrow 0$. This will completely suppress the usual bending of 5-branes, thereby freeing up one Coulomb branch direction per quiver node. We will also show how to account for the missing Higgs branch moduli in that setup.

The ordinary 5-brane web case corresponds to IIA at strong coupling, which, as we just mentioned, naturally has special unitary groups.

\section{A summary of the argument}

We will present our findings in a concise, sketchy way in this short section, in order to give the reader some intuition behind our line of thinking, and develop our detailed arguments in subsequent portions of this paper.

Usually, 5d SCFT's are described as UV fixed points for quiver gauge theories at the origins of their moduli spaces, and only semi-simple gauge groups are considered. Abelian factors are usually ruled out by the argument in \cite{Intriligator:1997pq, Seiberg:1996bd} that $U(1)$'s with matter will hit a Landau pole, or at the very least, will require a UV completion of some kind. 

The technology for building SCFT's in M-theory on non-compact, singular Calabi-Yau threefolds is an active research domain, initiated in \cite{Intriligator:1997pq}, but more recently studied in the references cited at the introduction among many more works. Most of these theories are understood by studying the singularity structure of the geometry, often relying upon F-theory experience.

There is a class of theories, however, that can be discussed in terms of type IIA perturbative string theory: M-theory on non-compact CY threefolds that admit a $\mathbb{C}^*$-fibration over a local K3. This includes infinitely many (but not all) toric threefolds.\footnote{A toric threefold does not necessarily give a Lagrangian theory.}

Based on works on three-dimensional theories \cite{Aganagic_2010, Benini:2009qs}, the works \cite{Closset:2018bjz,Saxena:2019wuy} developed a systematic method for reducing M-theory on a given toric threefold to type IIA on a local K3 with D6-branes. 
This picture makes it easy to read off the low-energy quiver gauge theory description associated with the singularity.

More generally, if a threefold is described as a $\mathbb{C}^*$-fibration over a local K3, then it is a simple matter to read off the gauge theory data from the IIA perspective.

An obvious question comes to mind: D6-branes naturally carry $U(N)$ gauge groups. How should this be accounted for? 
Although this issue is raised in the paper \cite{Closset:2018bjz}, we decided to give it more attention. We will make the following claim:
\vskip.5cm
\noindent{\bf First claim:} \emph{Type IIA string theory on a local K3 with an ALE metric, with D6-branes wrapping holomorphic compact curves gives rise to a quiver gauge theory with $U(N)$ gauge groups. We corroborate this by computing the dimensions of Higgs branches for various classes of theories.}
\vskip.5cm
    
This claim immediately raises the following question: Since a large portion of such IIA backgrounds could be obtained from 6d SCFT's built in IIB via reduction on a circle and T-duality, are we claiming that we are seeing inconsistent (anomalous) 6d SCFT's in IIB?
We answer this in the negative with the following claim:
\vskip.5cm    
    \noindent{\bf Second claim:} \emph{Type IIA string theory on a local K3 times a circle (transverse to the branes), on the other hand, confers a St\"uckelberg mass to all $U(1)$ factors of any quiver gauge theory.}
\vskip.5cm
The St\"uckelberg mechanism in question is triggered by the presence of the $C_5 \wedge F$ anomalous coupling on the D6-brane worldvolume theory.
Note, that this particular mechanism has been studied in a different context in F-theory in \cite{Grimm:2011tb} and in six-dimensional setups in \cite{Hanany:1997gh, Brunner:1997gf}.
Hence, whenever we try to relate these 5d theories to 6d, we are forced to have an extra circle that renders the $U(1)$ factors massive.

From the M-theory perspective, this means we are compactifying the $\cC$-fibers to elliptic fibers, which allows for an immediate interpretation in terms of F-theory. Once this transition is made, it can be shown directly in M-theory how $U(1)$ factors become massive: The corresponding normalizable two-forms become non-harmonic.

Finally, and most interestingly, comes the following question: It is well-known that the Higgs branch of these theories is realized in M-theory as the complex structure moduli space.\footnote{For non-compact singularities, one has to introduce an appropriate notion of normalizability in order to count genuinely 5d moduli.} Since the non-compact threefold gave us the Higgs branch for $U(N)$ quiver gauge theories, once we reduce on a circle (transverse to the branes) and cut the groups down to $SU(N)$, we expect extra Higgs branch directions to appear, since morally:\footnote{This formula is only valid if enough flavors are present.}
\begin{equation*}
    {\rm dim} \mathcal{M}_{\rm Higgs} = \# {\rm hypers} - \# {\rm vectors}\,.
\end{equation*}
The question is then: How are these extra moduli realized in IIA and M-theory? Our answer is the following:
\vskip.5cm
\noindent{\bf Third claim:} \emph{Upon compactifying on a transverse circle, the bulk closed string moduli of the K3 become five-dimensional, and hence dynamical. They come in exactly the right amounts to account for the enhanced Higgs branches}
\vskip.5cm
Note, that our second and third claim are linked. The extra bulk hypermultiplets are coupled to the Abelian factors via WZ terms, which allows for the St\"uckelberg mechanism.

In section \ref{sec:5branes}, we corroborate these claims in the dual IIB 5-brane setup. There, these mechanisms take a different form. Essentially, we find that, at weak $g_s^{IIA}$ coupling, T-duality will land us in IIB with $g_s \rightarrow 0$. In this regime, brane bending is suppressed, and this frees up one Coulomb branch modulus per quiver node, corroborating the unitary groups.
We will also corroborate the claim that compactifying on an extra circle will eliminate these $U(1)$'s again.

One puzzle does remain, however. The discussion in this paper pertains to perturbative type IIA string theory. Once we go to the strong string coupling limit, which translates to taking an ALE limit in M-theory (w.r.t. to the M-theory circle fibration), we do expect unitary groups to `lose' their $U(1)$ factors on their own, without requiring further compactification. 
As remarked in \cite{Closset:2018bjz}, this is unclear from a IIA perspective. Presumably, as $g_s \rightarrow \infty$, the D6-branes become increasingly delocalized, and their `center of mass' photons are no longer five-dimensional.

\section{Type IIA on local K3's with D6-branes} \label{sec:UNinIIA}

In this section, we describe our main string theory framework, from which we will be able to read off quiver gauge theory data at weak coupling.

We start with Type IIA string theory on the space $\rr^{1,5}\times S$, where $S$ is an hyperk\"ahler non-compact surface with compact two-spheres. For $S$ defined as the resolution of an orbifold $ \cc^2/\Gamma$, with $\Gamma \subset SU(2)$ a discrete group, the set of such two-cycles has a basis $\mathcal{C}_a$ ($a=1,...,r$) whose elements intersect each other like the Dynkin diagram of an ADE algebra of rank $r$. Here, we will assume an ALE metric. 

This background preserves sixteen supercharges. We call $x^0,...,x^5$ the coordinates on $\rr^{1,5}$ and $y^1,...,y^4$ the local coordinates on $S$.

The basis of 2-cycles $\mathcal{C}_a$ has a dual basis of normalizable two-forms $\alpha_a$ ($a=1,...,r$) on $S$. % (when $S$ is an ALF space, there is an extra normalizable two-form $\alpha_0$). 
Let us expand $C_3$ and $B_2$ as 
\be
C_3 = {\cal A}_M^a dx^M \wedge\alpha_a  + ...  \qquad\mbox{and}\qquad 
B_2 = b^a \alpha_a  + ...   \:.
\ee
The abelian gauge fields ${\cal A}_M^a$ ($M=0,...,5$) and the scalars $b_a$ propagate in six-dimensions.  The geometric moduli of the hyperk\"ahler metric  propagate in six-dimensions as well. They are given by the periods of the three hyperk\"ahler two-forms $\omega_i$ ($i=1,2,3$) on the curves $\mathcal{C}_a$'s. One can fix a K\"ahler structure by choosing the K\"ahler form to be, e.g., $J=\omega_3$ and the holomorphic $(2,0)$ form $\Omega_2=\omega_1+i\omega_2$; we call $\xi_a$ (real) the periods of $J$ and $\zeta_a$ (complex) the periods of $\Omega_2$.
The vector ${\cal A}_M^a$ and the scalars $b_a,\xi_a,\zeta_a$ sit together in a 6d $\Ntwo$ vector multiplet.

\

We now introduce a BPS D6-brane wrapping a compact curve $\mathcal{C}_{\bar a}$ in $S$ and extending along the directions $x^0,...,x^4$ inside $\rr^{1,5}$. This introduces a 5-dimensional $\mathcal{N}=1$ supersymmetric field theory with a vector multiplet $(A^{\bar a}_\mu, \phi_{\bar a})$ with $\mu=0,...,4$ ($\phi_{\bar a}$ parametrizes the motion of the D6-brane along $x^5$). 
The bulk $\Ntwo$ vector multiplet corresponding to this curve splits into a $\mathcal{N}=1$ vector multiplet $({\cal A}_\mu^{\bar a},\xi_{\bar a})$ and a hypermultiplet $(\zeta_{\bar a},b_{\bar a},{\cal A}_5^{\bar a})$. The fields in these multiplets propagate in six dimensions and are seen as non-dynamical background fields from the 5-dimensional D6-brane point of view. The modulus $\xi_{\bar a}$ controls the gauge coupling of the vector multiplet on the D6-brane.

\

One can generalize this situation, by wrapping $N_a$ D6-branes on the compact curves  $\mathcal{C}_a$
($a=1,...,r$). We  also allow D6-brane stacks on non-compact divisors of $S$; their fluctuations are non-dynamical in 5d, while their worldvolume gauge groups are flavor groups for the 5d theory.
The 5d spectrum is the following:
\begin{itemize}
\item one $\mathcal{N}=1$ vector multiplet in the adjoint representation of $U(N_a)$ for each compact curve $\mathcal{C}_a$;
\item one hypermultiplet in the bifundamental representation $({\bf N}_a,{\bf \bar{N}}_b)$ for each pair of intersecting compact curves $\mathcal{C}_a,\mathcal{C}_b$ (i.e. such that  $\mathcal{C}_a \cdot \mathcal{C}_b=1$);
\item one hypermultiplet in the fundamental representation ${\bf N}_a$ for each compact curve $\mathcal{C}_a$ that intersects a flavor brane. This hypermultiplet sits also in the fundamental representation of the corresponding flavor group.
\end{itemize}

This theory has a Coulomb branch and a Higgs branch (and mixed branches). We are interested in the last one.
The Higgs branch is obtained by giving vev's to all the scalars in the hypermultiplets. This breaks the gauge group to a subgroup (possibly $\{1\}$) by a Higgs mechanism. Some of the hypermultiplets are eaten by the  vector multiplets that become massive. The remaining neutral hypermultiplets are massless flat directions that parametrize the Higgs branch of the moduli space.

\section{M-theory uplift and moduli counting}\label{Sec:MthUplift}

In this section, we will uplift the type IIA setups described so far to M-theory on non-compact CY threefolds. These threefolds will be obtained by pairing up the M-theory circle with the $x_5$ transverse coordinate to make a $\cC$-fibration that collapses over the loci where D6-branes lie, following the general algebraic form\footnote{It is well-known that multi-Taub-NUT metrics admit a complex structure where they can be described in algebraic form \cite{Etesi:2008ew}.}
\be
u v = \Delta_{\rm D6}\,,
\ee
where $\Delta_{\rm D6}=0$ is the full D6-locus. These threefolds will be $\cC$-fibrations over resolved ALE spaces. We will keep the ALE bases resolved, in order to have weakly coupled quiver gauge theories.

Once we have such a description, we will count the complex structure moduli that deform the singularities localized along 5d subspaces. We will see that this counting gives the right Higgs branch dimensions for $U(N)$ quivers.

\subsection{Toric threefolds: The $A_n$ series} \label{sec:a-series}

In this section, we will show how to uplift our type IIA setups on resolved $\mathbb{C}^2/\mathbb{Z}_{n+1}$ with D6-branes to non-compact toric threefolds. Once the geometries are defined, we will count the complex structure moduli deformations that correspond to normalizable (and hence 5-dimensional) degrees of freedom. These will be identified with flat directions along the Higgs branch in the 5d effective theory.

We will describe the M-theory threefold $X_3$ as a $\mathbb{C}^*$-fibration over the toric IIA hyperk\"ahler surface. We will use the algebraic description of the form $u v = \Delta$, whereby $\Delta$ is a function pure of the base coordinates, parametrizing the positions of the various D6-branes involved, and the $uv$ part describes the $\mathbb{C}^*$-fiber, which collapses over the branes.

Let us define the following ambient fourfold $\mathsf{A}_4$:

\begin{equation}
    \begin{array}{c c c c c c c c c}
        u & v & z_1 & e_1 & e_2 & e_3 &\ldots & e_{n} & z_2  \\ \hline
        0 & 0 & 1 & -2 & 1 & 0 & \ldots & 0& 0 \\ 
        0 & 0 & 0 & 1 & -2 & 1 & \ldots & 0& 0\\ 
        \\
        0 & 0 & 0 &0 &0 & 0 & \ldots  & -2  & 1
    \end{array} \label{toricdef}
\end{equation}
The $(u,v)$ coordinates define the fiber, and the rest define the IIA base of the $\mathbb{C}^*$-fibration. The $z_{1, 2}$ coordinates correspond to two non-compact curves. The $e_i$ coordinates correspond to the exceptional spheres. 

The geometry is such that adjacent coordinates in \eqref{toricdef} correspond to curves that intersect. Note, that here we only have pairwise intersections, which means that only linear quivers are possible.

The fact that $u$ and $v$ have weights zero under the various $\mathbb{C}^{*}$-actions ensures that we have a balanced quiver, and that we have Chern-Simons levels equal to zero.
This choice imposes the constraint that our discriminant $\Delta$ be itself a function of multi-degree zero. So let us find a basis of multi-degree zero monomials. It turns out there are three such generators, subject to a relation:
\begin{equation}\label{InvariantsHomQuiv}
    X \equiv z_1^{n+1} e_1^{n} e_2^{n-1
    } \ldots e_{n}\,; \quad Y \equiv e_1  \ldots e_{n}^{n} z_2^{n+1} \,; \quad Z \equiv z_1 e_1 e_2  \ldots e_{n} z_2\,.
\end{equation}

They satisfy the relation $X Y = Z^{n+1}$. Indeed, these invariants correspond to the blow-down map of the resolved IIA orbifold.

In light of this relation, we see that we need only consider discriminants of the form
\begin{equation}
    \Delta = X^k Z^l\,, \quad {\rm or} \quad Z^l  Y^m
\end{equation}
Clearly, both options give equivalent theories, so we'll focus on the first one.

Let us first tackle the cases with $k=0$ and $l=0$ separately, and then move on to the mixed cases.

\subsubsection{Homogeneous linear quivers}

\begin{figure}[ht!]
\centering
\includegraphics[scale=.35]{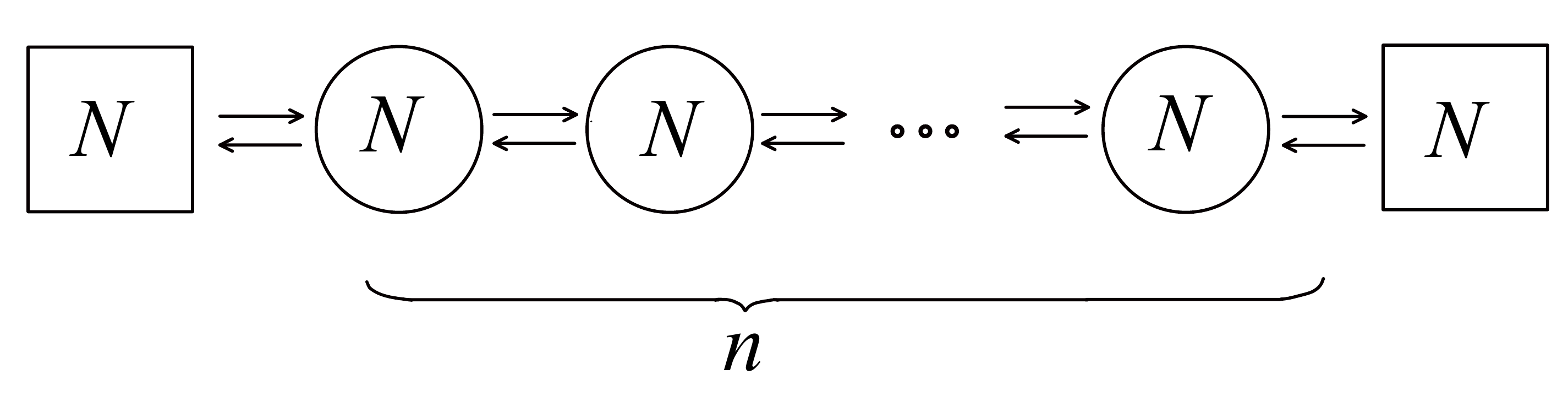}
\caption{Homogeneous linear quiver}
\label{HomQuiv}
\end{figure}

We will begin by studying `homogeneous' linear quivers despicted in Figure~\ref{HomQuiv}, meaning that all nodes in the quiver have the same rank (including the flavor nodes). The geometry in question is given by:
\begin{equation}
    P:= u v - Z^N = 0 \:,
\end{equation}
where $Z$ is defined in \eqref{InvariantsHomQuiv}. The $z_1^N$ and $z_2^N$ factors correspond to the two square nodes at the extremes of the quiver, and each $e_i^N$ factor represents a $U(N)$ node on this $i$-th exceptional sphere.

Let us now count possible deformations. A priori, one would compute the Jacobian ring:
\begin{equation}
    J = \mathbb{C}[u, v, z_1, z_2, e_i]/(dP) \cong \mathbb{C}[z_1, z_2, e_i]/(\tfrac{Z}{z_1}, \tfrac{Z}{z_2}, \tfrac{Z}{e_i})
\end{equation}
where the denominator runs over all $i=1, \ldots, n$. Counting the possibilities, we find an infinite number of possible deformations. So, clearly, if we are to compute the dimension of a Higgs branch, we must refine our search. 

The culprit here is the fact that the singularity is not isolated, but runs over the $z_1=0$ and $z_2=0$ divisors. This leads the infinitely many deformations. In physics terms, these extra modes correspond to vevs for the adjoint scalars living on the non-compact D6-branes. Since these extend over non-compact curves, they are not five-dimensional, but actually seven-dimensional. Hence, in the IR, we regard them as vevs for background (non-dynamical) fields, which appear as couplings in the effective theory.

How should we then isolate the truly five-dimensional modes? One way to go about it is to notice that our discriminant has the form:
\begin{equation}
    \Delta = z_1^N z_2^N  \left(\prod_i e_i\right)^N \:.
%    \Delta = z_1^N z_2^N  (\Pi(e))^N\,, \quad {\rm with} \quad \Pi(e) = \prod_i e_i
\end{equation}
If we have a deformation by a polynomial $\delta$ such that
\begin{equation}
    \Delta \mapsto (z_1^N z_2^N+\delta) \left(\prod_i e_i\right)^N\:,
\end{equation}
we should interpret this as a movement of the first or the second stack of flavor D6-branes, or a recombination of them. In any case, this will correspond to a vev for a seven-dimensional field.

Hence, from now on, we discard monomials proportional to $ \left(\prod_i e_i\right)^N$. We will see that this will not only cut the moduli space down to a finite-dimensional space, but will give the expected Higgs branch dimension on the nose.

We need to count terms of multi-degree zero w.r.t. all $n$ $\mathbb{C}^*$-actions defined in~\eqref{toricdef}. Hence, we can write everything in terms of our invariants $X, Y, Z$. Note that all three invariant coordinates are proportional to $\prod_i e_i$. Hence, we can immediately rule out anything in the ideal
\begin{equation}
(Z, X, Y)^N = \{X^{a} Y^{b} Z^{c}\, |\, a+b+c = N\}
\end{equation}
Keep in mind that products between $X$ and $Y$ are equivalent to powers of $Z$.
We will sort the terms by powers of $Z$. A generic deformation then looks as follows: 
\begin{equation}
    \delta = \sum_{i=0}^{N-2} Z^i\left(\sum_{j=1}^{N-i-1} (a_{i,j} X^j+b_{i,j} Y^j)\right)+\sum_{i=0}^{N-1}c_{i} Z^{i}\,.
\end{equation}
the number of parameters is then given by the sum
\begin{align}
 \#{\rm def}_\delta \,& = \, 2 \sum_{i=0}^{N-2} \left(\sum_{j=1}^{N-i-1} (1)\right)+\sum_{i=0}^{N-1} 1 =  2 \sum_{i=0}^{N-2} \left((N-i-1) \right)+N \\
 &=\, 2 \sum_{k=1}^{N-1} (1)+N = N^2
\end{align}

Hence, we deduce that the dimension of the \emph{normalizable} complex structure moduli space for this singularity is 
\begin{equation}
    {\rm dim}_{\mathbb{C}}\mathcal{M}_{c.s} = N^2\,.
\end{equation}
Note that this is independent of the order of the orbifold, but only depends on the number of branes in the picture.

The fact that we computed the \emph{complex} dimension to be $N^2$ stems from the fact the the complex structure moduli only give half of the Higgs branch. The other half is fibered over the former, and comes from $C_3$ axionic moduli.

Now let us compare this to gauge theory expectations. The general formula for the Higgs branch is:
\begin{equation}
    {\rm dim}_{\mathbb{H}} \mathcal{H} = \#{\rm hypers} - \#{\rm vectors}\,.
\end{equation}
For this quiver, there are $n+1$ edges, each contributing $N^2$ hypermultiplets, and $n$ nodes, each contributing $-N^2$ vectors. This adds up to
\begin{equation}
    {\rm dim}_{\mathbb{H}} \mathcal{H} = (n+1) N^2-(n) N^2 = N^2\,.
\end{equation}
It works.

\subsubsection{Ascending quivers}

\begin{figure}[ht!]
\centering
\includegraphics[scale=.35]{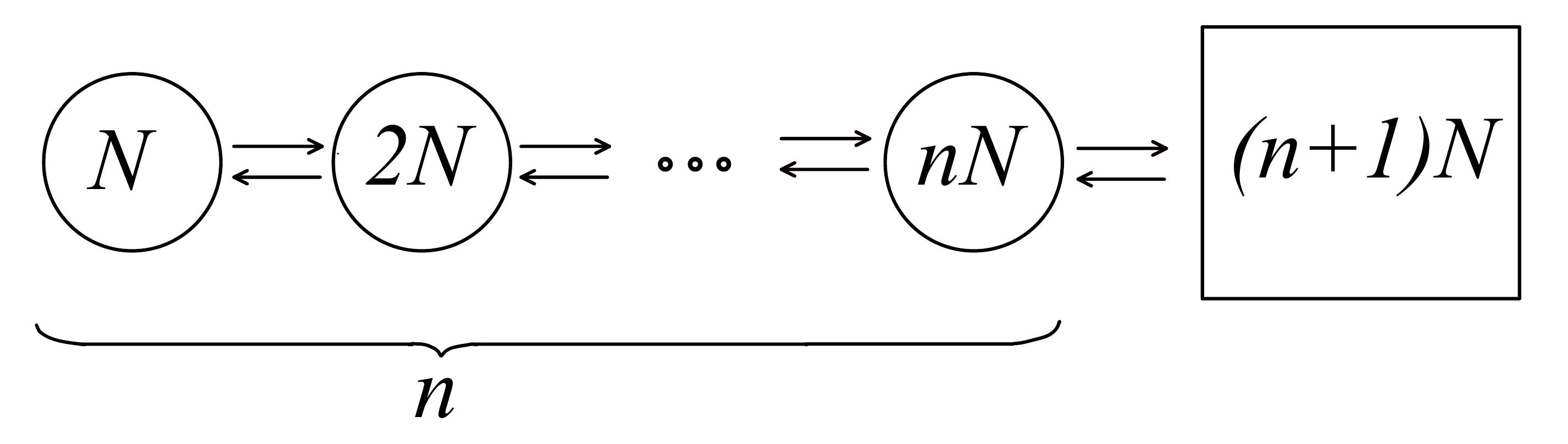}
\caption{Ascending quiver}
\label{AscQuiv}
\end{figure}

Now we move on to the next basic quiver, that is the generalization of the T$(SU)$ theories, but with the ranks all multiplied by a common factor of $N$. If we set $N=1$, this is the T$(SU(n+1))$ theory. We will refer to it as `ascending quiver', and it is depicted in Figure~\ref{AscQuiv}.

This quiver is realized in M-theory by the following hypersurface in the toric ambient space defined in \eqref{toricdef}:
\begin{equation}
    uv -\Delta=0\,, \quad {\rm for} \quad \Delta:= Y^N\,,
\end{equation}
with $Y$ defined in \eqref{InvariantsHomQuiv}. We can see that this discriminant distributed the branes in ascending fashion, starting with $N$ branes on the leftmost exceptional sphere, multipling the occupation number as we move to the right, ending with $(n+1)N$ flavor branes on the $z_2=0$ non-compact curve, where $n+1$ is the order of the orbifold.

Let us count the moduli for this problem. In terms of the original toric coordinates, this hypersurface is given by:
\begin{equation}
    uv = e_1^N e_2^{2N} \ldots e_n^{nN} z_2^{(n+1) N}\,.
\end{equation}
We want to exclude all deformations that correspond to movements of the non-compact stack. Hence, we are throwing out anything of the form
\begin{equation}
    e_1^N e_2^{2N} \ldots e_{n}^{n N}\rho \quad {\rm for \,\, some} \quad \rho\,.
\end{equation}
In terms of invariant coordinates, this means throwing out the ideal
\begin{equation}
    \{X^a Y^b Z^c\, |\, a+n b+c = n N\}\,.
\end{equation}
Let us arrange the calculation as follows: Since we are eliminating cross terms of the form $X^a Y^b$, we can split up the deformations as follows:
\begin{equation}
   \delta= \sum_{i=1}^{n N-1} X^i \left(\sum_{j=0}^{n N-1-i} a_{i, j} Z^j \right) + \sum_{i=1}^{N-1} Y^i \left(\sum_{j=0}^{n (N-i)-1} b_{i j} Z^j \right)+\sum_{j=0}^{n N-1} c_j Z^j\,.
\end{equation}
To count the terms, we set all variables and coefficients to one:
\begin{align}
     \#{\rm def}_\delta &= \sum_{i=1}^{n N-1}  \left(\sum_{j=0}^{n N-1-i} 1\right) + \sum_{i=1}^{N-1}  \left(\sum_{j=0}^{n (N-i)-1} 1\right )+\sum_{j=0}^{n N-1} 1 \\
    &=n N \big(N (n+1)-1\big)-\tfrac{1}{2} n N ( n N-1)- \tfrac{1}{2} n N (N-1)\\
    \\&=\tfrac{1}{2} n N^2 (n+1)\,,
\end{align}
where the last step only requires elementary manipulations.

Now let us see what the field theory side shows. Each link provides $kN (k+1)N$ hypermultiplets, where $k=1, \dots, n$. On the other hand, each node provides $(k N)^2$ vectors. Therefore, the dimension of the Higgs branch is given by:
\begin{equation}
     {\rm dim}_{\mathbb{H}} \mathcal{H} =N^2 \sum_{k=1}^{n} \left[k (k+1)-k^2\right]= \tfrac{1}{2} N^2 n (n+1)\,.
\end{equation}
A perfect match!

\subsection{Non-toric threefolds: The D and E series}

In this section we show some examples of threefolds that are not described as hypersurfaces in toric varieties, but as a complete intersection. Nevertheless we will be able to count the dimension of the Higgs branch, verifying that all the nodes of the quiver correspond to U(N) gauge groups.

\subsubsection{The $D_4$ quiver}

The $D_4$ singularity is described by the following equation in $\mathbb{C}^3$ with coordinates $X,Y,Z$.
\begin{equation}\label{D4singEq}
X^2 = Y\,Z\,(Y+Z) \:.
\end{equation}
The resolution of such singularity is a hypersurface in a toric threefold $\mathsf{A}_3$:
\begin{equation}
  \sigma \, w^2 =\lambda\,s\,t\,(s+t) \qquad \subset \qquad
    \begin{array}{c c c c c}
        s & t & w & \lambda & \sigma   \\ \hline
        1 & 1 & 1 & -1 & 0  \\ 
        0 & 0 & 1 & 1 & -1 \\ 
    \end{array} \label{toricdefD4}
\end{equation}

The four exceptional $\mathbb{P}^1$'s $\mathcal{C}_a$ ($a=1,...,4$) are given by
\begin{eqnarray}
\mathcal{C}_1 &:& \{ \,\sigma=0,\,s=0\,\}\\
\mathcal{C}_2 &:& \{ \,\sigma=0,\,t=0\,\}\\
\mathcal{C}_3 &:& \{ \,\sigma=0,\,s+t=0\,\}\\
\mathcal{C}_4 &:& \{ \,\sigma=0,\,\lambda=0\,\}
\end{eqnarray}

The blow-down map is determined by the invariant monomials (invariant under the two toric actions)
\begin{equation}\label{XYZD4}
X=w\,\sigma^2\lambda\,, \qquad  Y=s\,\sigma\,\lambda\,, \qquad  Z=t\,\sigma\,\lambda  \:,
\end{equation}
that verify the relation in \eqref{D4singEq}.

We now add the coordinates $u$ and $v$ to make a fivefold ambient space $\mathsf{A}_5$. As in the $A_n$ case, we choose their weights  under the toric $\mathbb{C}^\ast$ actions to be zero. The M-theory is again a $\mathbb{C}^\ast$ fibration over the resolved $D_4$ singularity of the form $uv=\Delta$, with $\Delta$ a polynomial of the toric coordinates $(w,s,t,\lambda,\sigma)$ with toric degrees zero. 

\begin{figure}[ht!]
\centering
\includegraphics[scale=.25]{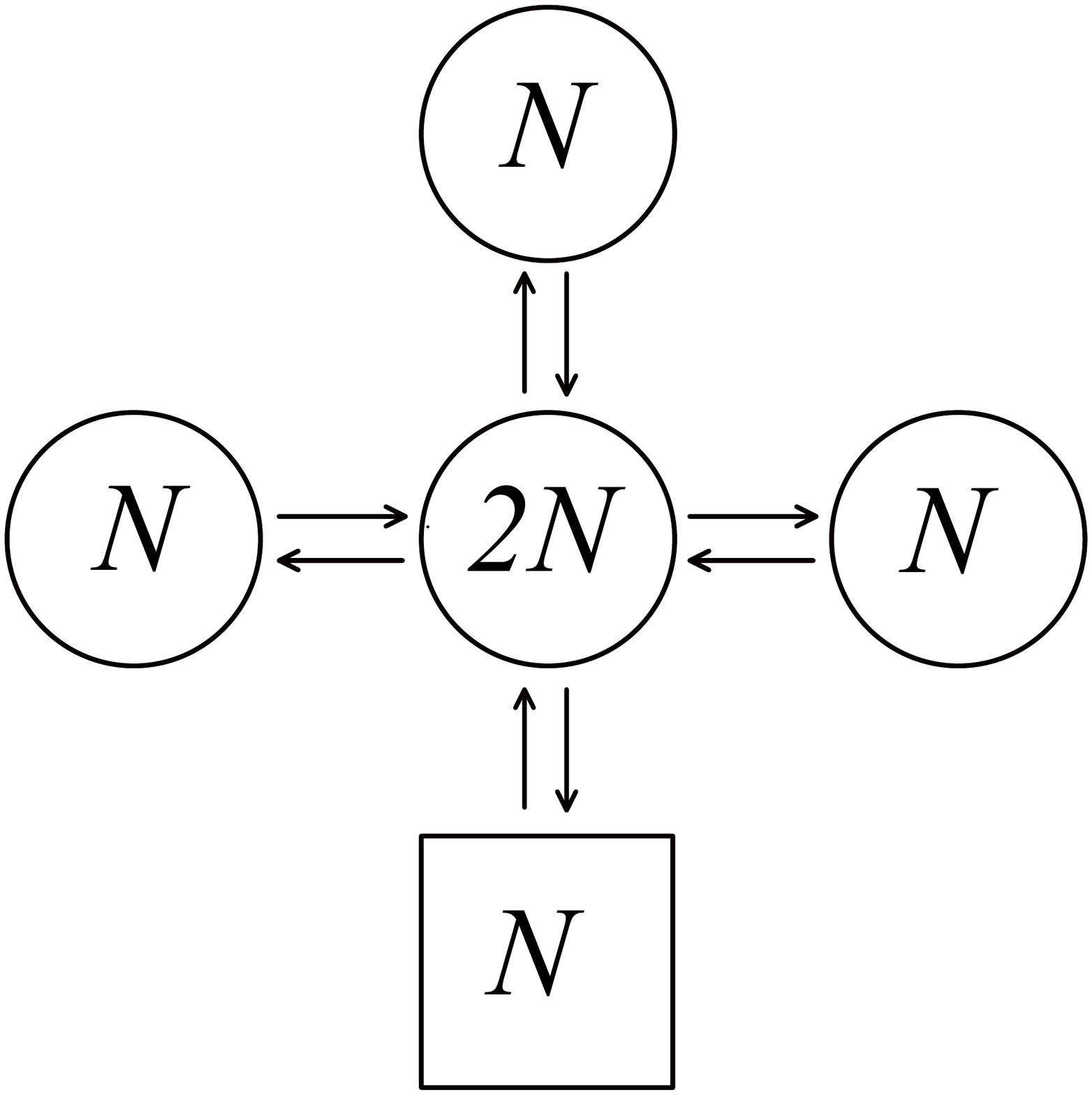}
\caption{$D_4$ quiver}
\label{D4quiv}
\end{figure}

For simplicity we study the case with $\Delta=(\gamma Y+\beta Z)^N$, with $\gamma,\beta\in\mathbb{C}$ (the most generic case is an straightforward generalization). The M-theory threefold is then given (using \eqref{XYZD4})  by the following complete intersection in $\mathsf{A}_5$:
\begin{equation}\label{X3D4eq}
\left\{
\begin{array}{l}
u\,v = \lambda^N\sigma^N\left( \gamma \,s+\beta \, t \right)^N\\
\sigma \, w^2 =\lambda\,s\,t\,(s+t)  \\
\end{array}
\right.
\end{equation}
We can read the type IIA brane configuration: we have $N$ D6-branes wrapping the curves $\mathcal{C}_1,\mathcal{C}_2,\mathcal{C}_3$ and $2N$ D6-branes wrapping $\mathcal{C}_4$; moreover we have $N$ D6-branes wrapping the non-compact curve given by $\gamma \,s+\beta \, t=0$ in the local K3 base.
The quiver of the corresponding field theory is reported in Figure~\ref{D4quiv}.

Let us count the 5d moduli. In order to exclude all deformations that correspond to movements of the non-compact stack, we throw out anything of degree zero that is proportional to $\lambda^N\sigma^N$. In other words, we keep the degree-zero monomials of the form
\begin{equation}
\lambda^m\sigma^{m+\ell}w^\ell P_{m-\ell}(s,t)     
%\lambda^{k+\ell}\sigma^{k+2\ell}w^\ell P_{k}(a,b)     
\end{equation}
with $P_{m-\ell}(s,t)$ a polynomial of degree $m-\ell$ in $s,t$, and such that $m\leq N-1$. Moreover,
we do not count the monomials with $\ell>1$, because one can use the second equation in \eqref{X3D4eq} to eliminate $\sigma^{k}w^{2k}$ in favour of powers of $\lambda,s,t$:
\begin{itemize}
\item  When  $\ell=0$, we have to count the monomials $\lambda^m\sigma^mP_m(s,t)$ with $m\leq N-1$: 
\begin{equation}
\# {\rm def}_0 = \sum_{m=0}^{N-1}(m+1)=\frac{N(N-1)}{2}+N=\frac{N(N+1)}{2} \:.
\end{equation}
\item  When $\ell=1$, we have to count the monomials $\lambda^m\sigma^{m+1}w\,P_{m-1}(s,t)$ with $m\leq N-1$: 
\begin{equation}
\# {\rm def}_1 = \sum_{m=1}^{N-1}m=\frac{N(N-1)}{2} \:.
\end{equation}
In total we have
\begin{equation}\label{numdefD4}
\# {\rm def} = \# {\rm def}_0 +\# {\rm def}_1 = \frac{N(N+1)}{2}+\frac{N(N-1)}{2}=N^2\:.
\end{equation}
\end{itemize}

We now check that it matches with the $D_4$ field theory with $U(N)^3\times U(2N)$ gauge group. We count $2N^2$ hypermultiplets for each link of the quiver, $N^2$ vector multiplets for each external gauge node and $4N^2$ vector multiplets for the internal node. Therefore, the dimension of the Higgs branch is given by:
\begin{equation}
     {\rm dim}_{\mathbb{H}} \mathcal{H} = 4\times 2N^2 - 3\times N^2 - 4N^2 = N^2\:.
\end{equation}
This matches with \eqref{numdefD4}.

\subsubsection{The $E_6$ quiver}

The $E_6$ singularity is described by the following equation in $\mathbb{C}^3$ with coordinates $X,Y,Z$.
\begin{equation}\label{E6singEq}
X^2 = Y^3 + Z^4 \:.
\end{equation}
The resolution of this singularity is a hypersurface in a toric threefold $\mathsf{A}_3$:
\begin{equation}
w^2 =v_1v_2^2v_1s^3+v_1^2t^4 \qquad \subset \qquad
    \begin{array}{c c c r r r r}
        s & t & w & v_1 & v_2 & v_3 & v_4   \\ \hline
        1 & 1 & 1 & -1 & 0 & 0 & 0  \\ 
        1 & 0 & 1 & 1 & -1 & 0 & 0  \\
        0 & 0 & 1 & 1 & 1 & -1 & 0  \\ 
        0 & 0 & 1 & 1 & 0 & 1 & -1  \\
    \end{array} \label{toricdefE6}
\end{equation}

The six exceptional $\mathbb{P}^1$ $\mathcal{C}_a$ ($a=1,...,6$) are given by
\begin{eqnarray}
\mathcal{C}_1 &:& \{ \,v_2=0,\,w+v_1t^2=0\,\}\\
\mathcal{C}_2 &:& \{ \,v_3=0,\,w+v_1t^2=0\,\}\\
\mathcal{C}_3 &:& \{ \,v_4=0,\,w^2-v_1v_2^2v_1s^3-v_1^2t^4=0\,\}\\
\mathcal{C}_4 &:& \{ \,v_3=0,\,w-v_1t^2=0\,\}\\
\mathcal{C}_5 &:& \{ \,v_2=0,\,w-v_1t^2=0\,\}\\
\mathcal{C}_6 &:& \{ \,v_1=0,\,w=0\,\}
\end{eqnarray}

The blow-down map is determined by the invariant monomials (invariant under the two toric actions)
\begin{equation}\label{XYZE6}
X=w\,v_1v_2^2v_3^4v_4^6\,, \qquad  Y=s\,v_1v_2^2v_3^3v_4^4\,, \qquad  Z=t\,v_1v_2v_3^2v_4^3  \:,
\end{equation}
that can be easily shown to verify \eqref{E6singEq}.

We  add the coordinates $u$ and $v$ to make a fivefold ambient space $\mathsf{A}_5$. As before they have zero degrees  under the toric $\mathbb{C}^\ast$ actions. The M-theory is again a $\mathbb{C}^\ast$ fibration over the resolved $E_6$ singularity of the form $uv=\Delta$, with $\Delta$ a polynomial of the toric coordinates $(w,s,t,v_1,...,v_4)$ with toric degrees zero. 

\begin{figure}[ht!]
\centering
\includegraphics[scale=.3]{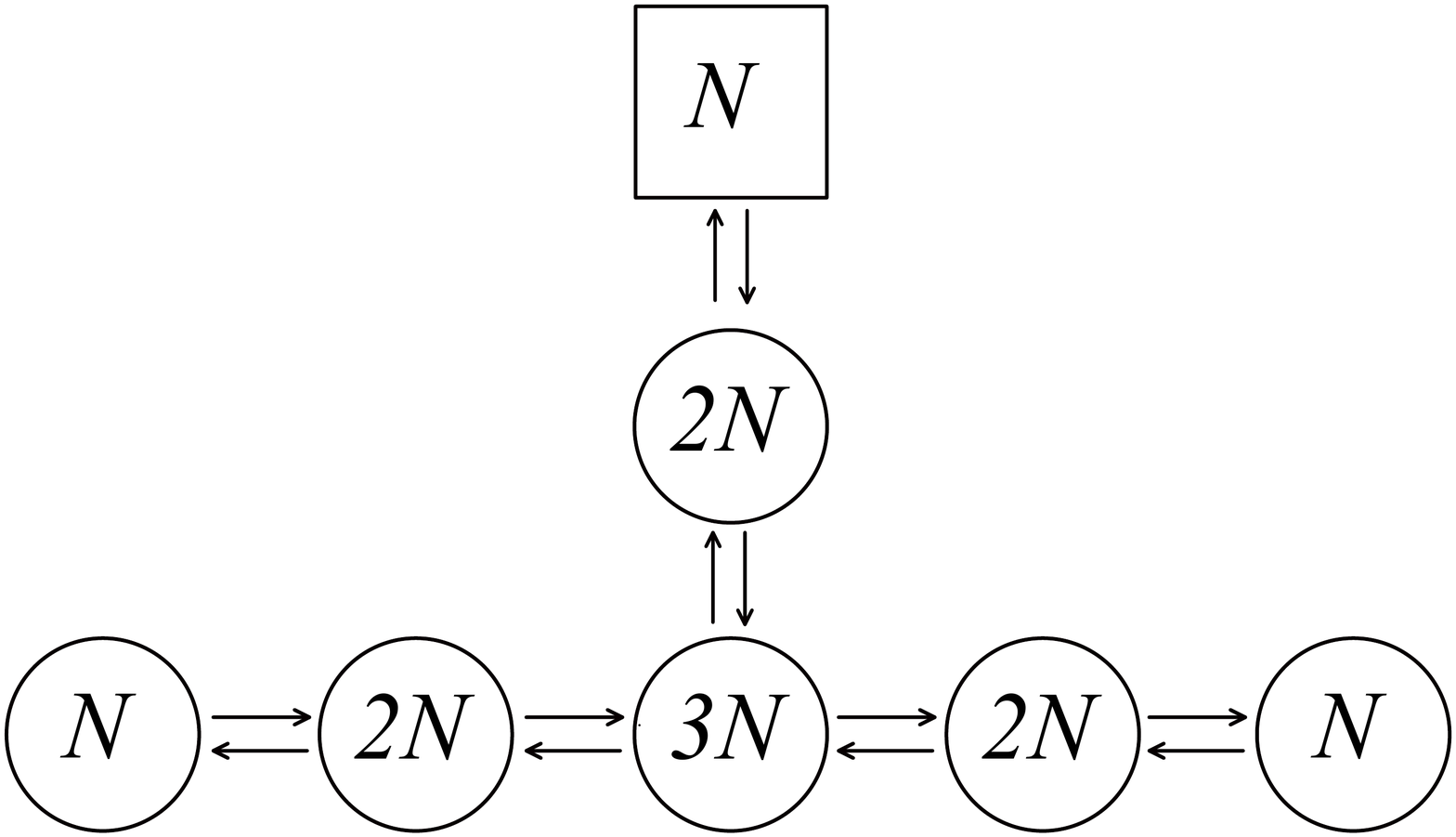}
\caption{$E_6$ quiver}
\label{fig:E6quiv}
\end{figure}

For simplicity we study the case with $\Delta=Z^N$ (the most generic case is a straightforward generalization). The M-theory threefold is then given (using \eqref{XYZE6})  by a complete intersection in $\mathsf{A}_5$:
\begin{equation}\label{X3E6eq}
\left\{
\begin{array}{l}
u\,v = v_1^Nv_2^Nv_3^{2N}v_4^{3N} b^N\\
%\sigma \, w^2 =\lambda\,a\,b\,(a+b)  
w^2 =v_1v_2^2v_1s^3+v_1^2t^4 \\
\end{array}
\right.
\end{equation}
We can read the type IIA brane configuration. First note that the locus $v_2=0$ is the union of the two curves $\mathcal{C}_1$ and $\mathcal{C}_5$; analogously, the locus $v_3=0$ is the union of the two curves $\mathcal{C}_2$ and $\mathcal{C}_4$; moreover the locus $v_1=0$ is twice the curve $\mathcal{C}_6$.
We then have $N\sigma_a$ D6-branes wrapping the curve $\mathcal{C}_a$, where $\sigma_a$ is the Dynkin label of the corresponding node. Moreover we have $N$ D6-branes wrapping the non-compact curve $t=0$.
The quiver of the corresponding field theory is reported in Figure~\ref{fig:E6quiv}.

Let us count the 5d moduli. In order to exclude all deformations that correspond to movements of the non-compact stack, we throw out anything of degree zero that is proportional to $ v_1^Nv_2^Nv_3^{2N}v_4^{3N}$. In other words, we keep the degree-zero monomials of the form
\begin{equation}
w^\ell(v_1v_2v_3^2v_4^3)^{n+m+\ell}s^mt^n v_2^{m+\ell} v_3^{m+2\ell} v_4^{m+3\ell}     
\end{equation}
with $n+m+\ell\leq N-1$. Moreover,
we do not count the monomials with $\ell>1$, because one can use the second equation in \eqref{X3E6eq} to eliminate $w^{2k}$ in favour of powers of $s,t,v_1,...,v_4$:
\begin{itemize}
\item  When  $\ell=0$, we have to count pairs $(n,m)$ such that $n+m=k$ and $k\leq N-1$; the number of pairs $(n,m)$ s.t. $n+m=k$ is equal to $k+1$, then
\begin{equation}
\# {\rm def}_0 = \sum_{k=0}^{N-1}(k+1)=\frac{N(N-1)}{2}+N=\frac{N(N+1)}{2} \:.
\end{equation}
\item  When $\ell=1$, we have $k\leq N-2$. Hence, 
\begin{equation}
\# {\rm def}_1 = \sum_{k=0}^{N-2}(k+1)=\frac{(N-1)(N-2)}{2} + N-1 =\frac{N(N-1)}{2}\:.
\end{equation}
In total we have
\begin{equation}
\# {\rm def} = \# {\rm def}_0 +\# {\rm def}_1 = \frac{N(N+1)}{2}+\frac{N(N-1)}{2}=N^2\:.
\end{equation}
\end{itemize}

Again, it matches with the corresponding $E_6$ field theory with $U(N)^2\times U(2N)^3\times U(3N)$ gauge group. We count $\sigma_a\sigma_bN^2$ hypermultiplets for each link of the quiver (again $\sigma_a$ is the Dynkin label of the node $a$), $2N^2$ hypermultiplets at the link with the non-compact curve, and $(\sigma_aN)^2$ vector multiplet for each  gauge node. Therefore, the dimension of the Higgs branch is given by:
\begin{equation}
     {\rm dim}_{\mathbb{H}} \mathcal{H} = 2N^2 \times 3 +6N^2 \times 3  - N^2\times 2 -(2N)^2	\times 3 -(3N)^2  = N^2\:.
\end{equation}

\subsection{Underbalanced quivers}
So far, we have only considered so-called \emph{balanced} quivers, meaning each node has $N_f = 2 N_c$. Those theories have the property that they can be completely Higgsed. From the IIA perspective, it means that the compact and non-compact D6-branes can be recombined into non-compact D6-branes, and completely move off the exceptional $\pp^1$'s.

In this section, we will consider simple examples where the number of flavors drops below the balancing threshold. The consequence is that such theories retain a residual gauge group, which makes the field theory analysis slightly more nuanced. This translates into residual rigid compact D6-branes that cannot escape the exceptional $\pp^1$'s.

In the following we  consider a single $U(N)$ factor in the gauge group. The corresponding geometry is a $\mathbb{C}^\ast$ fibration over a resolved $A_1$ singularity. It is given by the equation $u\, v = \Delta$ in the ambient space 
\begin{equation}
    \begin{array}{c c c c c}
        u & v & z_1 & e &  z_2  \\ \hline
        N_f-2N & 0 & 1 & -2 &  1 \\ 
    \end{array} \label{toricdefunbal}
\end{equation}
where we choose the D6-brane locus to be $\Delta=e^N z_1^{N_f}$, without any loss of generality.\footnote{In principle one could choose any polynomial of degree $N_f$ in $z_1$ and $z_2$, instead of $z_1^{N_f}$, but the presented results would not change.}
We distinguish the cases where $N_f$ is even or odd.

\subsubsection*{$U(N)$ with $N_f=2M$}

The threefold is given by $u\,v = e^N z_1^{2M}$, with $M< N$ (the case with $M=N$ has been studied in Section~\ref{sec:a-series}).

We now count the 5d moduli. We should include deformations of the hypersurface equations, i.e. of degree $-2N+2M$, that are not proportional to $e^N$, in order to exclude the deformations  corresponding to movements of the non-compact stack. The most generic deformation is then
\begin{equation}\label{Eq:Unbal2Mflav}
  u\,v = e^N z_1^{2M} + \sum_{j=1}^{M} e^{N-j} P_{2(M-j)}(z_1,z_2) \:,
\end{equation}
where $P_m(z_1,z_2)$ are polynomials of degree $m$ in $z_1$ and $z_2$. Each of them contributes with $m+1$ deformations. The total number of moduli is then
\begin{equation}
\# {\rm def}_{N_f=2M} = \sum_{j=1}^M (2M-2j+1) = M(2M+1)-2\sum_{j=1}^Mj=M^2 \:.
\end{equation}
From equation \eqref{Eq:Unbal2Mflav} we can also read the unbroken gauge group with the charged spectrum. In fact, we can rewrite the deformed D6-locus as
\begin{equation}
\Delta^{\rm def} = e^{N-M} \left[  e^Mz_1^{2M} + \sum_{j=1}^{M} e^{M-j} P_{2(M-j)}(z_1,z_2) \right]\:.
\end{equation}
After recombination one still finds $N-M$ D6-branes wrapping the compact $\pp^1$. Moreover, one sees that the surviving gauge group has no flavor as the recombined D6-brane does not intersect the exceptional $\pp^1$ (the intersection would be at $e=0$ and $P_0=0$, but $P_0$ is a non-zero constant).

Let us compare this with the gauge field expectations. 
Giving non-zero vev to the hypermultiplet scalars, one breaks the gauge group to a $U(K)$ subgroup of $U(N)$. 
The Adjoint and the fundamental representations of $U(N)$ breaks as
\begin{eqnarray}
  \mathbf{N^2} &\rightarrow& \mathbf{K^2} \oplus (N-K) \times \left( \mathbf{K} \oplus \mathbf{\bar{K}}\right) \oplus (N-K)^2\times \mathbf{1} \\
    \mathbf{N} &\rightarrow& \mathbf{K} \oplus (N-K) \times  \mathbf{1} 
\end{eqnarray}
The massive  gauge bosons need to acquire mass from a suitable component of the hypermultiplets; in particular each vector multiplet that becomes massive needs to eat one hypermultiplet. We can then count the part of the $2M$ hypermultiplets in the $\mathbf{N}$ of $U(N)$ that is eaten:
\begin{itemize}
\item The singlet massive vector multiplets eat $(N-K)^2$ singlet hypermultiplets. We are then left with $2M(N-K) - (N-K)^2$ neutral hypermultiplets.

\item Each of the $2(N-K)$ massive vector multiplets in the $\mathbf{K}$  of $U(K)$ eats one hypermultiplet in the same representation. The remaining number of  charged hypermultiplets is then $2M - 2(N-K)$.
\end{itemize}
Since the last number cannot be negative, it is clear that the minimum gauge group we can obtain is $U(K)$ with $K=N-M$ and with no charged hypermultiplets. The number of neutral hypermutliplet, i.e. the dimension of the Higgs branch, is $2M(N-K) - (N-K)^2|_{K=N-M} = M^2$, that is in agreement with our geometric computation. Notice that the geometric setup gives also the correct surviving gauge group and charged spectrum.

\subsubsection*{$U(N)$ with $N_f=2M+1$}

The threefold is now given by $u\,v = e^N z_1^{2M+1}$, with $M< N$.

Again we consider the deformations that are not proportional to $e^N$. The deformed threefold is 
\begin{equation}\label{Eq:Unbal2Mp1flav}
  u\,v = e^N z_1^{2M+1} + \sum_{j=1}^{M} e^{N-j} P_{2(M-j)+1}(z_1,z_2)  \:.
\end{equation}
The total number of moduli is then
\begin{equation}
\# {\rm def}_{N_f=2M+1} = \sum_{j=1}^M (2M-2j+2) = M(2M+2)-2\sum_{j=1}^Mj=M(M+1) \:.
\end{equation}
To explicitly read the unbroken gauge group with the charged spectrum we  rewrite the deformed D6-locus as
\begin{equation}
\Delta^{\rm def} = e^{N-M} \left[  e^Mz_1^{2M} + \sum_{j=1}^{M} e^{M-j} P_{2(M-j)+1}(z_1,z_2) \right]\:.
\end{equation}
After recombination one  finds $N-M$ D6-branes wrapping the compact $\pp^1$. There is still one flavor, as the recombined D6-brane intersect the gauge stack at one point (the intersection is at $e=0$ and $P_1(z_1,z_2)=0$).

Let us compare this with the gauge field expectation. 
The breaking pattern is analogous as before. However, in the present case we begin with one more flavor.
The number of remaining charged hypermultiplets is now equal to $2M+1-2(N-K)$. Hence again the maximal breaking will occur for $K=N-M$, but now a massless hypermultiplet in the fundamental representation of $U(N-M)$ remains after higgsing. The number of neutral hypermultiplets, i.e. the dimension of the Higgs branch is $(2M+1)(N-K) - (N-K)^2|_{K=N-M} = M(M+1)$. Again we have a perfect match with our geometric computation both with respect to the neutral hypermultiplets and the spectrum of the surviving gauge group.\footnote{This argument does not directly apply to $U(1)$ with $N_f=1$. However, it is easy to see from a 4d perspective that  F-term and D-term equations imply no Higgs mechanism. This is confirmed by the geometric computation, as $\Delta=e\,z_1$ admits no deformations.}

\section{SU(N) gauge theories}

We will now present a perturbative IIA mechanism that will turn any unitary quiver into a special unitary one via a St\"uckelberg mechanism.

We consider the same setup as above, but we now compactify the direction $x^5$ on the circle $S^1_{x^5}$. Nothing changes on the D6-brane worldvolume, as the D6-branes are points on the 5th direction. However, after compactifying $x^5$, the bulk fields have zero modes propagating in 5d. These fields now enter as dynamical fields in the 5d theory. Hence, the 5d spectrum is the same as in Section~\ref{sec:UNinIIA} with the addition of the (bulk) neutral hypermultiplets $(\zeta_{a},b_{a}+i{\cal A}_5^{a})$
and the (bulk) decoupled\footnote{Nothing is charged under the $U(1)$ corresponding to  the vector multiplet $({\cal A}_\mu^{a},\xi_{ a})$ (unless a sphere $\mathcal{C}_a$ collapses).} vector multiplets $({\cal A}_\mu^{a},\xi_{a})$.

Due to a St\"uckelberg coupling of the bulk fields with the D6-brane, the overall $U(1)$ vector multiplet on each D6-brane stack gets a mass term, as we now explain. 

The 7-dimensional D6-brane action includes the following coupling
\be\label{StuckD6Coup}
\int_{D6} F\wedge \iota^* C_5
\ee
where $C_5$ is the dual of the RR threeform $C_3$ and $\iota^*$ is the pull-back on the D6-brane worldvolume. One can expand $C_5$ as
\be 
C_5 = c_{(3)MNP}^a dx^M\wedge dx^N\wedge dx^P \wedge\alpha_a  + ... 
\ee
where $c^a_{(3)}$ are five-dimensional three-forms, dual to the scalars ${\cal A}_5^a$ coming from $C_3$. 

Let us consider a D6-brane wrapping the curve $\mathcal{C}_{\bar a}$. The action term \eqref{StuckD6Coup} reduces in 5d to
\begin{equation}
 \int_{\mathbb{R}^{1,4}}  F \wedge c^{\bar a}_{(3)}
\end{equation}
where we used that $\int_{\mathcal{C}_{\bar a}}\alpha_a = \delta^{\bar a}_a$. This drives a St\"uckelberg mechanism in five dimensions: The vector becomes massive eating the scalar ${\cal A}_5^a$. Since $\mathcal{N}=1$ supersymmetry must be preserved in 5d, all fields in the vector multiplet must become massive, i.e. we have the so-called super-Higgs mechanism: The vector multiplet $(A^{\bar a}_\mu, \phi_{\bar a})$ {\it eats} the hypermultiplet $(\zeta_{\bar a},b_{\bar a}+i{\cal A}_5^{\bar a})$, making a massive (long) vector multiplet.  %\footnote{The vector multiplet $({\cal A}_\mu^{\bar a},\xi_{\bar a})$ remains in the spectrum, but no \emph{massless} matter is charged under the corresponding $U(1)$ (unless the sphere $\mathcal{C}_{\bar a}$ collapses).} 

The St\"uckelberg mechanism requires the eaten scalar to be a dynamical field. This happens only when we compactify the $x^5$ direction. 

When the dust settles, and all the hypermultiplets have zero vev, the remaining 5d massless spectrum is the following:
\begin{itemize}
\item one $\mathcal{N}=1$ vector multiplet in the adjoint representation of $SU(N_a)$ for each compact curve $\mathcal{C}_a$;
\item one hypermultiplet in the bifundamental representation $({\bf N}_a,{\bf \bar{N}}_b)$ for each pair of intersecting compact curves $\mathcal{C}_a,\mathcal{C}_b$ (i.e. such that  $\mathcal{C}_a \cdot \mathcal{C}_b=1$);
\item one hypermultiplet in the fundamental representation ${\bf N}_a$ for each compact curve $\mathcal{C}_a$ that intersects a flavor brane. This hypermultiplet sits also in the fundamental representation of the corresponding flavor group.
\end{itemize}

The 5d theory again has Coulomb and Higgs branches. However, we should now see a bigger Higgs branch than before, since we know that:
\be
{\rm dim} \mathcal{M}_{\rm Higgs} = \#{\rm hypers} - \#{\rm vectors}\,,
\ee
and we have just lost a bunch of vector multiplets. So how are these extra hypermultiplets realized in this IIA setup?

The answer is that, we have already included $r$  (bulk) hypermultiplets that now get eaten by the Abelian factors. This frees up $r$ charged (open string) hypermultiplets, which can now develop non-trivial vev's, thereby enhancing the dimension of the Higgs branch by~$r$.

%We now need to include the $r$ neutral (bulk) hypermultiplets. When one give non-zero vev to all the hypermultiplets, the gauge group is broken due to the vev of the charged hypers. In the setup we are considering, one has also the St\"uckelberg coupling. The starting gauge group is again $U(N)$ for each stack of brane; the number of eatable hypers is however increased by one for each stack when the $x^5$ direction is compactified. When the hyper vevs break the overall $U(1)$ of the stack wrapping $\mathcal{C}_a$, the hyper eaten by the $U(1)$ vector multiplet will be a combination of the bulk hyper and the charged hypers. In the situation when $x^5\in \mathbb{R}$, only and charged hypers were present and were eaten. Hence when an overall $U(1)$ is broken by the charged hyper vevs, the Higgs branch includes an extra hyper related to the corresponding curve.

We have then the following relation between the dimension $\dim \mathcal{H}_{SU}$ of the Higgs branch when $x^5\in S^1$ and the dimension $\dim \mathcal{H}_{U}$ of the Higgs branch when $x^5\in\mathbb{R}$.
\be\label{dimHBSUU}
\dim \mathcal{H}_{SU} =  \dim \mathcal{H}_{U} + k \:,
\ee
where $k$ is the number of overall $U(1)$'s that would be broken along the Higgs branch $ \mathcal{H}_{U}$ of the theory with unitary groups.\footnote{In the SU-case, these U(1)'s eat a combination of the bulk and the open string hypermultiplets, leaving one more combination as a flat direction.}

One can understand this also from the D6-brane point of view: Consider for simplicity a stack of $N$ D6-branes on two intersecting curves $\mathcal{C}_a,\,\mathcal{C}_b$. The gauge group is $SU(N)\times SU(N)$ and there is a hypermultiplet in the $({\bf N},{\bf \bar{N}})$ representation. %The two bulk hypers are eaten by the two overall $U(1)$'s of the corresponding D6-brane stacks. In particular, t
The periods of the holomorphic (2,0)-form over the two curves (these moduli sit in the hypermultiplets eaten in the St\"uckelberg mechanism) are fixed to be zero by the BPS condition on the D6-branes: this is a signal that they are no longer allowed deformations.
When we give a vev to the charged hypermultiplet, the two stacks recombine in a stack wrapping a curve $\mathcal{C}$ in the homology class $[\mathcal{C}] = [\mathcal{C}_a]+[\mathcal{C}_b]$. Only this curve is require to be holomorphic by the BPS condition, letting free one combination of the two periods.
%Now the gauge group is the diagonal $SU(N)$, and the overall $U(1)$ vector multiplet eats the combination of the bulk hyper relative to the curve $\mathcal{C}$, while the other combination is left massless. In particular, t
%The BPS condition now fixes only one combination of the periods $\zeta_a,\zeta_b$, while the other combination is a modulus.  
In this simple case, the Higgs branch dimension is then equal to one, as it should be for the Higgs branch of $SU(N)\times SU(N)$ with one bifundamental (the residual gauge group is $SU(N)$, given by the recombined D6-brane).

\section{5d from M-theory on elliptically fibered CY$_3$}

In type IIA we have considered what happens by compactifying the $x^5$ direction, which lies transverse to the D6-branes and the local K3. This compactification is what triggered the St\"uckelberg mechanism.
In the dual M-theory setup studied in the previous section, this direction lives along the $\mathbb{C}^\ast$-fiber of the CY threefold. Remember that the generic $\mathbb{C}^\ast$-fiber is a cylinder. Compactifying the $\mathbb{R}$ direction, produces an elliptic curve that is then fibered over the base of the original $\mathbb{C}^\ast$-fibration.

In this section, we will study the consistency conditions to achieve such a compactification, and show how to implement it.

%\subsection{M-theory on elliptically fibered threefolds: Consistency condition}
\subsection{Consistency condition for compactifying the $\mathbb{C}^\ast$-fiber}

In this section, we will study the conditions for compactifying the real $x^5$-direction  transverse to both the D6-branes and the local K3. From the IIA point of view, we have to remember that D6-branes backreact on the metric of the transverse space, and it is not guaranteed that putting $x^5$ on a circle is possible. 

In the absence of D6-branes, we start out with a product of $S \times \mathbb{R}$, whereby $S$ has a constant metric along the real line. However, the presence of D6-branes will induce a piecewise linear $x^5$-dependence of the K\"ahler parameters of $S$. This precludes a compactification to a circle, unless further branes are added to the picture.

In order to understand this piecewise linear dependence, we will start from the M-theory geometry, which automatically takes the backreaction into account, and reduce to IIA via symplectic reduction. In the purely toric case, this is explained in detail in \cite{Closset:2018bjz}. However, we will proceed in a slightly different way that will lend itself to generalization beyond the toric case, since we are also interested in local K3 surfaces of D and E type.

The type IIA setup on $\mathbb{R}^{1,5}\times S$ with D6-branes wrapping curves of $S$ is dual to M-theory on a threefold $X_3$, which is the $\mathbb{C}^\ast$ fibration over $S$ described in Section~\ref{Sec:MthUplift}. This has the form
\begin{equation}
u v = \Delta 
\end{equation}
where $\Delta$ is the polynomial on $S$ whose zeroes give the location of the D6-branes. Depending on the setup, $u, v$, and $\Delta$ are sections of appropriate line bundles over $S$. 

The $\mathbb{C}^*$-fiber over each point given by $uv = $constant, can be decomposed into $\mathbb{R} \times S^1$. The circle action is given by 
\be \label{circleaction}
U(1): (u, v) \mapsto (e^{i \theta} u, e^{-i \theta} v)\,,
\ee 
and each orbit is an $S^1$ that we interpret as the M-theory circle. The real direction can be identified via the symplectic reduction method, and we find that $x^5$ corresponds to the Hamiltonian function:
\be \label{momentmap}
x^5 := |u|^2-|v|^2\,.
\ee
A sufficient condition to guarantee that the projection:
\be
\pi_{M/IIA}: X_3 \longrightarrow S \times \mathbb{R}_{x^5}
\ee
actually yields a trivial product $S \times \mathbb{R}$, is to make sure that $u, v$ and $\Delta$ are sections of the trivial line bundle over S. 

This implies that $x_5$, as expressed by the moment map in \eqref{momentmap}, will be a section of a trivial real line bundle over $S$. Moreover, we will see that, with this condition, the K\"ahler volumes of the compact curves in $S$ will have constant values along the $x^5$ coordinate, so that we can periodically identify that direction. However, this might be overkill, and we would like to determine the necessary conditions.

Let us zoom in on a single compact sphere $\mathcal{C} \subset S$. Locally, $S$ can be regarded as the total space of the normal bundle $N_{\mathcal{C} \subset S}$, which is $\mathcal{O}(-2)_{\mathbb{P}^1_{\mathcal{C}}}$. 

We will be very pedantic in our description of this space, in order to set the stage for our subsequent M-theory/IIA reduction. Let us begin describe this IIA space in the absence of branes. We can represent this torically as the non-compact weighted projective space:
\be
\{\mathbb{C}^{3}-(z_1, z_2, e) \neq (0,0, e)\}/\mathbb{C}^*
\ee
where the three complex coordinates have the following homogeneous weights under the $\mathbb{C}^*$-action:
\be
\begin{array}{c c c}
z_1 & z_2 & e\\
1 & 1 & -2
\end{array}\,, \quad {\rm with} \quad (z_1, z_2, e) \neq (0,0, e)\,.
\ee 
This is the definition of this toric space as a holomorphic quotient by a $\mathbb{C}^*$ action. It is well-known that this same quotient can be recast in the language of the symplectic reduction, by decomposing $\mathbb{C}^* \cong \mathbb{R} \times U(1)$ into scale a phase actions. The scale action is fixed by a so-called ``D-term condition'' (in $d=4, \mathcal{N}=1$ language):
\be
|z_1|^2+|z_2|^2-2 |e|^2 = \xi\,,
\ee
where $\xi$, the so-called ``FI constant'', is a real constant that measures the volume of the $\mathbb{P}^1$ under the K\"ahler form $J$:
\be
\xi := \int_{\{e=0\}} J \:.
\ee
Here, $e=0$ corresponds to the $\mathbb{P}^1$, and the $z_i$ to the normal directions. Now that the $\mathbb{R}$-action is broken by this relation, we simply quotient by the remaining $U(1)$-action.

Now we would like to add branes to the picture and uplift this to M-theory. Assuming that there are $N_c$ branes on the compact curve, and $N_f$ branes on non-compact curves, we can describe the local M-theory geometry as the hypersurface:
\be \label{branelift}
u v = e^{N_c} P^{(N_f)}(z_1, z_2)
\ee
inside the following ambient space
\be
\begin{array}{c c c c c}
u & v & z_1 & z_2 & e\\
q_u & q_v & 1 & 1 & -2
\end{array}
\ee
where, for consistency, $q_u+q_v = N_f- 2 N_c$. The ``D-term condition'' is now
\be
q_u |u|^2+q_v |v|^2+|z_1|^2+|z_2|^2-2 |e|^2 = \xi_M\,.
\ee
This is the local model for our CY threefold on which we define M-theory.

In order to reduce this to IIA, we reduce with respect to the circle action on $u$ and $v$ defined in \eqref{circleaction}. We then have the following description of the ambient space as a quotient by $U(1)^2$ acting as follows:
\be
\begin{array}{c c c c c}
u & v & z_1 & z_2 & e\\
q_u & q_v & 1 & 1 & -2\\
1 & -1 & 0 & 0 & 0
\end{array}
\ee
with the same hypersurface condition as before, and two ``D-term conditions'':
\begin{align} \label{mthdterm}
q_u |u|^2+q_v |v|^2+|z_1|^2+|z_2|^2-2 |e|^2 &= \xi_M\,, \\
|u|^2-|v|^2 &= x^5\,,
\end{align}
where we have conveniently named the second ``FI constant'' $x^5$. 

Let us now restrict to the curve $e=0$, which sets the r.h.s. of \eqref{branelift} to zero. For $x^5 \neq 0$, depending on its sign, the second ``D-term condition'' in \eqref{mthdterm} will forbid either $u$ or $v$ from vanishing, and equation \eqref{branelift} will force the other variable to vanish. Let us choose without loss of generality $x^5>0$, such that $u \neq 0$ and $v=0$. In that case, we are left with the following description of the family of curves $\mathcal{C}$ over $x^5$:
\be
\begin{array}{c c c}
u & z_1 & z_2 \\
q_u & 1 & 1 \\
1 & 0 & 0 
\end{array}
\ee
with 
\begin{align}
q_u |u|^2+|z_1|^2+|z_2|^2 &= \xi_M-q_u x^5\,, \\
|u|^2 &= x^5\,.
\end{align}
It is convenient to take a linear combination of the actions and ``D-term conditions'' such that $u$ no longer mixes with the $z_i$:
\be
\begin{array}{c c c}
u & z_1 & z_2 \\
0 & 1 & 1 \\
1 & 0 & 0 
\end{array}
\ee
with 
\begin{align}
|z_1|^2+|z_2|^2 &= \xi_M\,, \\
|u|^2 &= x^5\,.
\end{align}
Now we see that we can safely eliminate $u$, since we can fix its phase and norm, and we are simply left with a $\mathbb{P}^1$ with K\"ahler volume:
\be
\int_{{\mathcal{C}}} J = \xi_M - q_u x^5
\ee
The case $x^5<0$ is treated similarly, swapping the roles of $u$ and $v$ (and minding a sign). We can therefore regard this as a local K3 fibered over the $x^5$-line, with an effective running K\"ahler parameter whose period is
\be
\int_{\mathcal{C}} J_{\rm eff}= \begin{cases}  \xi_M-q_u x^5   \,,\quad x^5 >0\\  \xi_M+q_v x^5    \,, \quad x^5 <0\end{cases}\,.
\ee
In order to compactify $x^5$, we need this volume to be periodic or constant, therefore, we must have $q_u = q_v = 0$. Note that this implies $N_f = 2 N_c$, but it is not implied by it. 

This condition can now be applied to each curve individually, regardless of the ADE type of local K3 at hand. The condition amounts to imposing that we consider only so-called \emph{balanced quivers} without Chern Simons terms, since the effective CS levels are given by \cite{Benini:2009qs}:
\be
k_{eff} = \tfrac{1}{2} \left. \frac{d \left(\int_{\mathcal{C}} J_{\rm eff} \right)}{d x_5}\right|^{x_5 = 0^+}_{x_5 = 0^-} = \tfrac{1}{2}(q_v-q_u)\,.
\ee

\subsection{The decompactification limit in M-theory}\label{Sec:MthLimit}

We now describe the limit one needs to take on the elliptic fibration to go back to the $\mathbb{C}^\ast$ fibration, such that the two backgrounds are dual to the type IIA with $x^5$ living respectively on a circle and on $\mathbb{R}$.

Let us consider M-theory on $T^2$ with radii $R_{11}$ and $R_5$ (the first is the M-theory circle and the second is the $x^5$ direction), i.e. with volume $v=R_{11}R_5$ and complex structure $\tau=\tau_1+i\tau_2$ where $\tau_2=R_{11}/R_5$.
The eleven dimensional metric is ($x^{11}$ is the M-theory circle direction):
\begin{equation}
 ds^2_M=\frac{v}{\tau_2}\left((dx^{11}+\tau_1d\hat{x}^5)^2+\tau_2^2(d\hat{x}^5)^2\right)+ds_9^2
\end{equation}
where $\hat{x}^5$ has periodicity $2\pi$.
On the other hand, the relation between the M-theory metric $ds^2_M$ and the type IIA metric $ds^2_{IIA}$ is 
\begin{equation}
 ds^2_M=e^{4\varphi/3}(dx+C_1)^2+e^{-2\varphi/3}ds_{IIA}^2
\end{equation}
From this we read immediately that $e^{2\varphi/3}=\sqrt{v/\tau_2}=R_{11}$, as expected, and the type IIA metric is
\begin{equation}
ds_{IIA}^2 = e^{2\varphi/3} \left(  (dx^5)^2 + ds_9^2\right)
\end{equation}
where we have rescaled $x^5=R_5\hat{x}^5$ so that $x^5$ has periodicity $2\pi R_5$.

The decompactification limit in type IIA is $R_5\rightarrow \infty$ keeping $\varphi$ fixed. This corresponds on the M-theory side to take $R_5\rightarrow \infty$ keeping $R_{11}$ fixed, i.e. 
\begin{equation}
\tau\rightarrow i\,\infty \,,\qquad v\rightarrow \infty \qquad\qquad\mbox{with}\qquad v/\tau_2 \,\,\,\mbox{fixed}
\end{equation}
In particular, in terms of the $j$-function $j(\tau)$ of the elliptic curve, this means $j\rightarrow\infty$.

\subsection{Elliptic threefolds: Algebro-geometric description}

We now describe the elliptic fibration algebraically, in a form that goes to the wanted $\mathbb{C}^\ast$ fibration in the limit described in Section~\ref{Sec:MthLimit}.
The $\mathbb{C}^\ast$ fibration $uv=\Delta$  can be equivalently written as 
\begin{equation}
y^2 = x^2 +  \Delta 
\end{equation}
by a simple change of coordinates: $u=y-x$, $v=y+x$.

Our claim is that the elliptic threefold that uplifts the type IIA setup when $x^5$ is compactified is algebraically described by
\begin{equation}\label{EllFibx2}
y^2 = \varepsilon x^3 +  x^2 +  \Delta 
\end{equation}
with decompactification limit $\varepsilon\rightarrow 0$. In fact, this equation describes an elliptic fibration that degenerates over the discriminant of the cubic on the right hand side, i.e. over\footnote{
One can bring \eqref{EllFibx2} in the canonical Weierstrass form $y^2=X^3+f\,X+g$, by redefining $x=\frac{X}{\varepsilon^{1/3}} - \frac{1}{3 \varepsilon}$. One obtains $y^2=X^3-\frac{1}{3\varepsilon^{4/3}}X+\Delta+\frac{2}{27\varepsilon^2}$.
One can then use the known formula for the discriminant and the function $j=\frac{4(24f)^3}{4f^3+27g^2}$. 
}
\begin{equation}
\Delta_{ell.fib.} = -\Delta \left( 4+27\varepsilon^2\Delta  \right)\:.
\end{equation}
Moreover the $j$-function is
\begin{equation}
j=\frac{2048}{\varepsilon^2\Delta\left( 4+27\epsilon^2\Delta  \right)}
\end{equation}
where we can immediately check that $\epsilon\rightarrow 0$ sends $j\rightarrow \infty$, as required for the decompactification limit.

We notice that the elliptic fiber degenerates where the D6-brane sit (i.e. at $\Delta=0$); there is also an extra component of the discriminant, that introduces neither new gauge fields nor extra matter (the two loci do not intersect each other) and that disappears to infinity in the $\Delta$-plane when one takes the decompactification limit.

\

The Calabi-Yau threefold  has now a {\it compact} fiber and a non-compact hyperk\"ahler surface base. The complex structure moduli of the base are given by deformations of the holomorphic (2,0) form $\Omega_2$. 
Since the fiber directions are compact, the periods of $\Omega_2$ on the compact curves of $S$ provide complex scalar fields propagating in 5d.\footnote{This happens because of the existence of normalizable harmonic two-forms corresponding to the compact 2-cycles, analogously with the type IIA case.}

The elliptic fibration (as well as the $\mathbb{C}^\ast$-fibration) obstructs some of these complex deformations of $S$. In particular the locus where the fiber degenerates must be a holomorphic curve on the base $S$, i.e. the periods of the holomorphic (2,0)-form of $S$ are forced to be zero on such a curve. In our case we have the constraint that $\Omega_2$ integrated over (the components of) the locus $\Delta=0$ must vanish. 

The deformations of the threefold studied in the Section~\ref{Sec:MthUplift} are deformations of the discriminant locus $\Delta=0$. In the cases we studied, the most generic deformation makes the degeneration locus a connected non-compact curve (in homology class $[\Delta]$) that does not intersect any compact curve. This provides no obstruction for the periods of $\Omega_2$ on the compact curves $\mathcal{C}_a$.
%If the locus $\Delta=0$ wrap a non-compact divisor in $S$ with zero intersections with the compact curves $\mathcal{C}_a$, then this gives no obstruction. 
Hence, all the $r$ complex structure moduli $\zeta_a$ of $S$ must be included in the counting of the (5d) complex structures of the CY threefold, and the dimension of the Higgs branch is increased by $r$ units with respect to what happend for the $\mathbb{C}^\ast$-fibration. This also matches with formula \eqref{dimHBSUU}.
%: the fact that we can recombine all the compact and non-compact D6-branes means that there are enough charged hypermultiplets to break all the $r$ $U(1)$ gauge groups (hence $k=r$ in \eqref{dimHBSUU}.
%\todo{Are there situations in which $k<r$? i.e. $[\Delta]$ intersects some compact curve and correspondingly U(1)s are not all broken by charged hypers?}

\subsection{Massive U(1)'s in M-theory on elliptically fibered threefolds}

In this section we explain why we do not have abelian gauge symmetries when we consider an elliptic fibration instead of a $\mathbb{C}^\ast$ fibration. This phenomenon was discovered and explained in the F-theory literature by \cite{Grimm:2011tb}, and later elucidated from a geometric standpoint in \cite{Braun_2014}.

\subsubsection{Supergravity explanation}
Let us recall the argument put forward by \cite{Grimm:2011tb} to understand how the St\"uckelberg mechanism explained from the IIA perspective manifests in M-theory.

Normally, in order to obtain a photon from M-theory, we would look for a normalizable, closed two-form $\omega$ in the threefold $X_3$, and take the following KK Ansatz for the supergravity $C_3$ form
\be
C_3 = \omega \wedge A \quad {\rm where} \quad d \omega = 0\,,
\ee
where $A$ is the photon in the low energy theory. The reduction of the kinetic term will then give us:
\be
\int_{X_3 \times \mathbb{R}^{1,4}} d C_3 \wedge \ast d C_3 \sim \int_{\mathbb{R}^{1,4}} dA \wedge \ast dA \:.
\ee

Take instead a two-form $\tilde \omega$ that is not closed, and choose the following Ansatz:
\be
C_3 = \tilde \omega \wedge A+ \varphi \chi_3\,, \quad {\rm where} \quad d\tilde \omega = \chi_3\,,
\ee
where $\varphi$ is a low energy scalar. Now the kinetic term becomes
\be
\int_{X_3 \times \mathbb{R}^{1,4}} d C_3 \wedge \ast d C_3 \sim \int_{\mathbb{R}^{1,4}} d A \wedge \ast dA + (A+d \varphi) \wedge \ast (A+d \varphi)\,.
\ee
Hence, the St\"uckelberg mechanism arises from the non-closure (or non-harmonicity) of the two-form.

\subsubsection{Algebro-geometric explanation}
In this section, we will translate the non-closure of the normalizable two-form of the previous section into a statement about non-K\"ahlerity of small resolutions.

We will consider a simple case that exemplifies all the relevant issues, i.e. $U(1)$ gauge theory with two charged hypermultiplets. This can be realized either by a homogeneous linear quiver or by an ascending quiver. Let us consider the former for simplicity. The $\mathbb{C}^\ast$ fibration is given by
\begin{equation}
uv=z_1e\,z_2   \qquad\qquad\mbox{or equivalently}\qquad\qquad y^2=x^2+z_1e\,z_2\:,
\end{equation}
where $(z_1,e,z_2)$ have toric weights $(1,-2,1)$ and where $e=0$ is the exceptional $\mathbb{P}^1$ (this is a particular case of \eqref{toricdef}); the $u,v$ and $x,y$ coordinates are related by $u=y-x$ and~$v=y+x$.

This space has two conifold singularities,  at $(x,y,z_1,e)$ and at $(x,y,e,z_2)$. Correspondingly there are three pairs of non-Cartier divisors $D_{z_1}^\pm=(x\pm y,z_1)$,  $D_{e}^\pm=(x\pm y,e)$ and $D_{z_2}^\pm=(x\pm y,z_2)$. 

The conifold singularities can be resolved. The M2-branes wrapping the exceptional curves give rise to two hypermultiplets. The normalizable {\it harmonic} taub-nut two-form are in one-to-one correspondence with the divisors $\alpha_{z_1}=D_{z_1}^+-D_{z_1}^-$, $\alpha_{e}=D_{e}^+-D_{e}^-$ and $\alpha_{z_2}=D_{z_2}^+-D_{z_2}^-$. Each harmonic two-forms give rise to a $U(1)$ vector multiplet, under which the hypermultiplets are charged. The three vector multiplets propagate respectively along $z_1=0$, $e=0$ and $z_2=0$. Hence only the second one is a gauge boson propagating effectively in 5d, while the other two are seen as background vector multiplets in the 5d theory (as they propagate in 7d).

This spectrum (one $U(1)$ vector multiplet and two charged hypermultiplets) matches with the expectation in type IIA, where we have one D6-brane wrapping the disconnected locus $z_1e\,z_2=0$, i.e. three D6-branes, one of which wrapping a compact curve, and two intersection loci giving rise to charged hypermultiplets.

\

Let us now compactify the $x^5$ direction. The elliptic fibration in now described by
\begin{equation}
y^2=\varepsilon x^3 +  x^2+z_1e\,z_2\:.
\end{equation}
This equation can be rearranged in the following form
\begin{equation}
 y^2 - z_1e\,z_2 = x^2 (\varepsilon x + 1)\:,
\end{equation}
i.e. it is two deformed $A_1$ singularities fibered over the  $x$-plane: when the r.h.s. vanishes, the l.h.s. describes a surface with two $A_1$ singularities, while at generic $x$ these singularities are deformed. This manifold is singular at $(x,y,z_1,e)$ and at $(x,y,e,z_1)$. As explained in detail in \cite{Braun_2014} (Section 4.2), this manifold admits only a non-K\"ahler resolution. The normalizable two-form $\alpha_e$ in the $\mathbb{C}^\ast$ fibration is now non-harmonic and the corresponding $U(1)$ gauge field has then a KK mass term \cite{Grimm:2011tb}. 

This explains in the M-theory setup why the compactification of the $x^5$ direction removes the abelian gauge bosons from the spectrum.\footnote{We also check that the Higgs branch dimension is increased by $r=1$ unit, passing from the $\mathbb{C}^\ast$ fibration to the elliptic fibration. A generic connected curve in the class  $[\Delta]$ does not intersect the compact curve $e=0$, hence the fibration does not obstruct the periods of $\Omega_2$ on $\mathcal{C}_e$.}

\section{5-brane web perspective}\label{sec:5branes}
Having studied the type IIA and M-theory embeddings of unitary quivers, let us now turn to the IIB 5-brane web perspective, in order to gain further insight into these $U(1)$'s.

We will actually study these webs in the limit $g_s^{IIB} \rightarrow 0$. In this regime, the main points will be the following:
\begin{enumerate}
\item The brane webs no longer behave as webs, but will behave exactly as Hanany-Witten setups in 3d \cite{Hanany:1996ie}. No brane bending takes place, hence everything will be consistent with $U(n)$ gauge groups. The enhancement of the Coulomb branch is then visible as up and down movements of D5-segments between parallel NS5-branes and $(1,1)$ 5-branes.\footnote{More precisely, although branes will no longer bend, charge conservation will still impose that when a D5 and an NS5 meet, a $(1,1)$ 5-brane will emerge from the intersection.}

\item When we compactify one direction longitudinal to the NS5's but transverse to the D5's, the `baryonic' branches become manifestly dynamical. They correspond to NS5 movements off the $(p,q)$-plane, as usual, only now the NS5's are truly 5d objects.
\end{enumerate}

\subsection{Unitary quivers from 5-brane webs}
Let us first define our conventions for standard 5-brane webs in  Table~\ref{tab:5branesusual}.
\begin{table}[h!]
\be
\begin{tabular}{c|cccccccccc}
 & 0 & 1 & 2 & 3 & 4 & 5 & 6 & 7 & 8 & 9\\
D5 & --- & --- & --- & --- & ---  &   &  --- & &  &  \\
 NS5 & --- & --- & --- & --- & ---  &   --- &    &   & &  \\ 
(1,1) 5-branes & --- & --- & --- & --- & ---     &    \multicolumn{2}{c}{angle} & & & \\ 
 7-branes & --- & --- & --- & --- & ---  &   && ---    &  --- & ---\\
 &  &  &  &  &   &  && && $S^1$ 
 \end{tabular}\nonumber
 \ee
 \caption{Brane setup.}
 \label{tab:5branesusual}
 \end{table}
The relation between this setup and our IIA setup is as follows: Starting with IIA on an $A_N$-type ALF space, we wrap all D6-branes (compact or non-compact) along the $S^1$-fiber of the ALF. After T-duality, they become D5-branes. At the same time, this non-trivially fibered $S^1$ becomes the $x^9$ direction in Table~\ref{tab:5branesusual}. The fixed points of the fibration T-dualize to NS5-branes that are pointlike in the $x^9$-circle, as indicated. The 7-branes are ingredients that will be brought in later. % in the discussion.
 
In order to set the stage for the discussion, let us remind the reader of the relations between the various couplings in string theory. Take M-theory on a torus of radii $(R_{11}, R_{9})$. The various 10d couplings are related as follows:
\begin{align}
\lambda_A &= R_{11}^{3/2}  & \lambda_B &= R_{11}/R_{9}\\
r_A^{st} &=R_9 \lambda_A^{1/3}= 	R_9	\sqrt{R_{11}}\; &\qquad \quad  r_B^{st}& \sim 1/r_A^{st} = 1/(R_9 \sqrt{R_{11}}) \nonumber
%r_A^E&=r_A^{st} \lambda_A^{-1/4}= R_9 R_{11}^{1/8}&\qquad  r_B^E&=r_B^{st}\lambda_B^{-1/4} =1/(R_9^{5/4} R_{11}^{3/4})\,.
\end{align}
Here, $\lambda_{A}$ and $r_A^{st}$ are the type IIA string coupling and radius of the IIA circle in string frame, respectively; the second column shows the IIB counterparts.\footnote{In Einstein frame the radia are $r_A^E=r_A^{st} \lambda_A^{-1/4}= R_9 R_{11}^{1/8}$ and $r_B^E=r_B^{st}\lambda_B^{-1/4} =1/(R_9^{5/4} R_{11}^{3/4})$.} Here, $r_B$ represents the radius of the $x^9$-direction in Table~\ref{tab:5branesusual}.

As a preview, instead of taking the familiar vanishing area limit of the M-theory torus (aka the \emph{F-theory limit}), we will keep $R_{11}$ finite, and send $R_9$ to infinity, so that at finite $\lambda_A$ we will have $\lambda_B \rightarrow 0$. 

\begin{figure}[ht!]
\centering
\includegraphics[scale=.7]{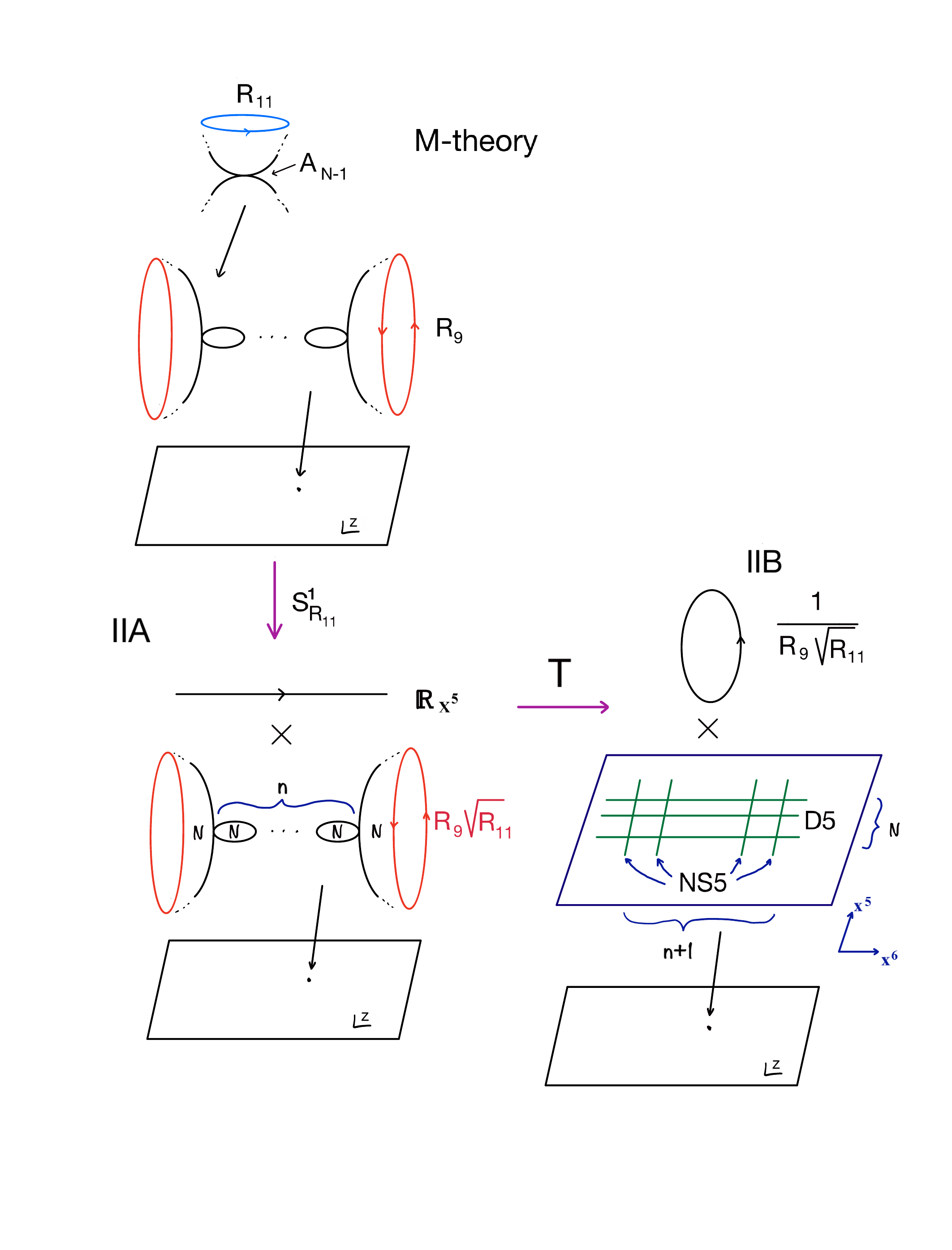} 
\caption{IIA on an ALF space. (The $z$-plane is spanned by the coordinates $x^7,x^8$, while the $\mathbb{C}^\ast$-fiber of the ALF space is spanned by $x^6$ and the circle coordinate $x^9$.)}
\label{fig:finite_radius}
\end{figure}

Let us get more specific. Take M-theory on a $\cC$-fibration over a local K3. The local K3 is itself a multi-centered Taub-NUT (ALF) with asymptotic radius $R_9$, and the $\cC$-fibration contains the ``M-theory circle'' with asymptotic radius $R_{11}$. This sets the IIA string coupling to $R_{11}^{3/2}$ as summarized above. The setup is summarized below, and in Figure~\ref{fig:finite_radius}:
\tikzstyle{block} = [draw,rectangle,thick,minimum height=2em,minimum width=2em] \tikzstyle{line} = [draw, very thick, color=black!50, -latex']
\begin{center}
\begin{tikzpicture}[node distance = 2cm, auto] 
    \node [block] (M) {M-theory on $\cC$-fibration over ALF}; 
    \node [block, below of=M] (IIA) {IIA on ALF with asymptotic radius $R_9 \sqrt{R_{11}}$}; 
    \node [block, right=30pt of IIA] (IIB) {IIB on $S^1$ with radius $1/(R_9 \sqrt{R_{11}})$};
    \path [line] (M) -- (IIA); 
 \path [line] (IIA) -- (IIB);
\end{tikzpicture}
\end{center}

In type IIA, the asymptotic radius of the multi-centered TN is not $R_9$, but gets rescaled to $R_9 \sqrt{R_{11}}$. After a T-duality, the TN geometry becomes flat space times a circle of radius $1/(R_9 \sqrt{R_{11}})$, with some NS5-branes, and the D6-branes dualize to D5-branes.\footnote{This happens for this particular arrangement. If we choose the non-compact D6-branes in generic positions, we will get D7-branes.}

\begin{figure}[ht!]
\centering
\includegraphics[scale=.5]{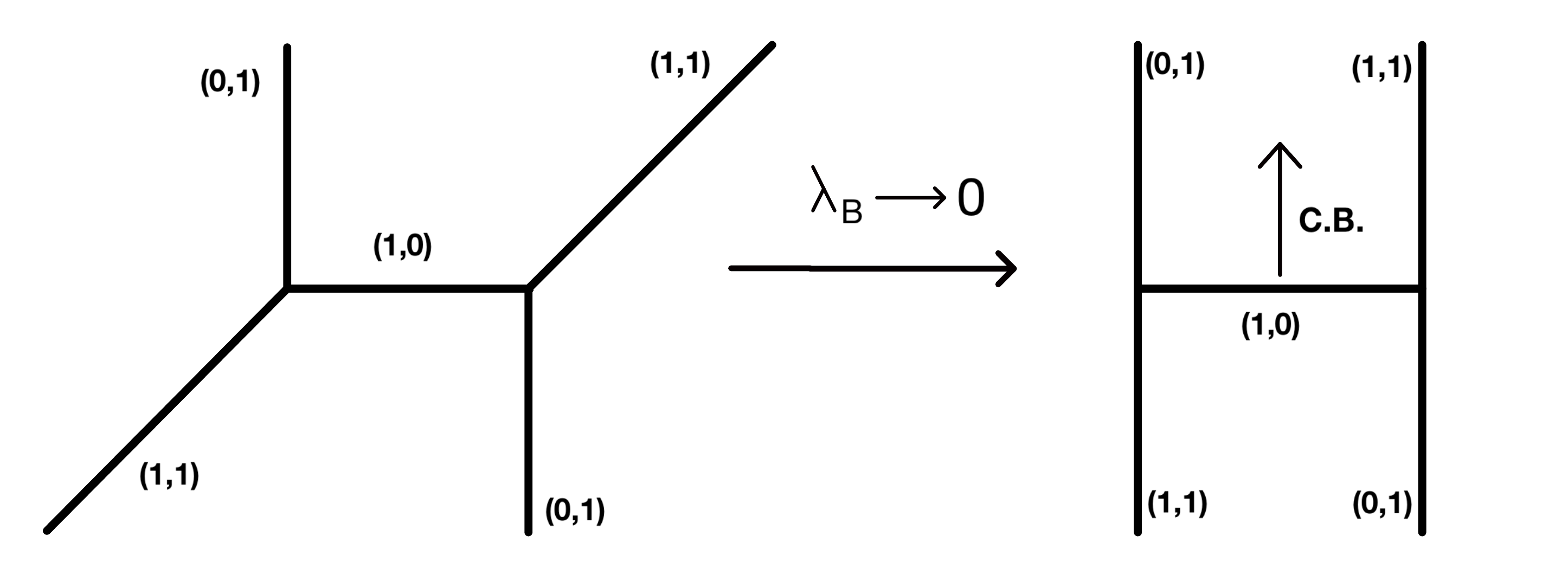} 
\caption{Branes don't bend at $\lambda_B=0$. A $U(1)$ photon becomes dynamical. The Coulomb Branch is represented by moving the D5-segment up and down.}
\label{fig:nobending}
\end{figure}

Now consider taking finite IIA string coupling, and sending the asymptotic TN radius to infinity. This lands us in perturbative IIA on an ALE space (i.e. a resolved orbifold of $\cc^2$). On the IIB side, we have the radius completely shrinking, and the string coupling going to zero. 
\be
\lambda_A < \infty,\, R_9 \rightarrow \infty \qquad \Rightarrow \qquad \lambda_B \rightarrow 0 \,.
\ee
The upshot is that now, the rules for 5-brane webs get modified. The slope-condition for a $(p, q)$-brane on the $(x^6, x^5)$-plane (depicted as a parallelogram on the r.h.s. in Figure~\ref{fig:finite_radius}) is the following:
\be
(\Delta x^6: \Delta x^5) =(p:q/\lambda_B).
\ee
Sending $\lambda_B \rightarrow 0$, we thus find that all $(p, q)$-branes with $q \neq 0$ will be vertical, and D5-branes will be horizontal. Hence, if a D5-brane meets an NS5 and a $(1,1)$ (by charge conservation), the latter two will together form a straight vertical line. This is depicted in Figure~\ref{fig:nobending}. Now the situation is analogous to 3d Hanany-Witten setups: D3-segments suspended between parallel NS5-branes are free to move up and down the diagram, representing a real direction of the Coulomb branch associated with overall $U(1)$ factors on unitary quivers. 

Hence, we can draw the conclusion that we have 5d unitary quivers in this regime. 

\begin{figure}[ht!]
\centering
\includegraphics[scale=.6]{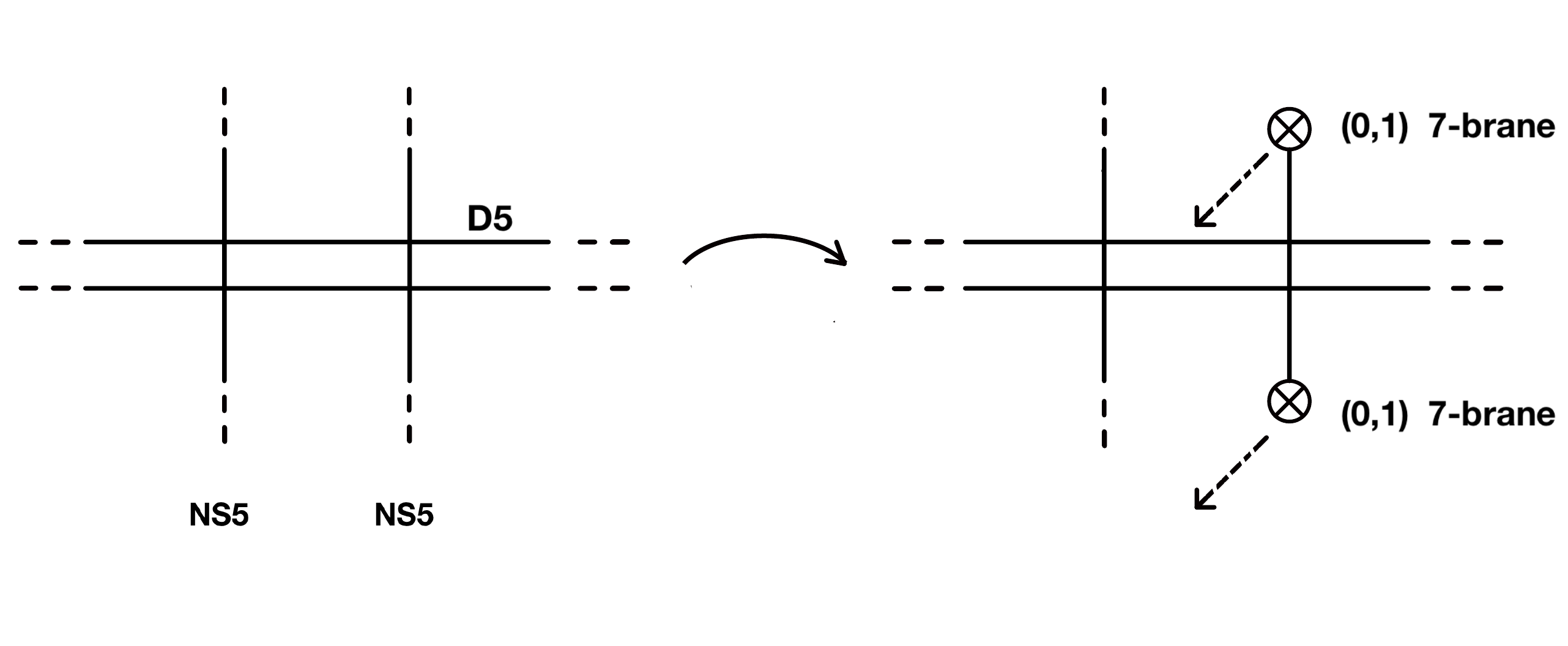} 
\caption{$SU(2)$ with $N_f=4$, moving onto a `baryonic branch'.}
\label{fig:sevenbranes}
\end{figure}

An important check to run here is the structure of the Higgs branch. More specifically, do we find a baryonic branch in these setups or not? If the answer is affirmative, then we cannot have a unitary group. Let us recall how baryonic branches are visualized in the 5-brane web framework when we have, say a single gauge node with $SU(N_c)$ with $N_f \geq N_c$:

The color branes are realized by $N_c$ finite suspended D5-segments, and the flavors by $N_f$ semi-infinite D5-branes. If we have at least $N_c$ semi-infinite branes all on one side of the color branes, then we align them with the latter, and reconnect. This frees up an NS$5$-brane that can move in the directions transverse to the $(p, q)$-plane, realizing the baryonic branch. However, in order to interpret this as a vev of a dynamical field, this NS5 must itself be suspended between two $(0,1)$-sevenbranes, in order to make its worldvolume truly five-dimensional. Otherwise, this would just be a ``global deformation'', corresponding to a parameter of the theory. This is illustrated in Figure~\ref{fig:sevenbranes}.

So at this point we should ask: Do we still see a baryonic branch here? The answer, interestingly, is ``no''. The $(0,1)$-sevenbranes are incompatible with the weak coupling limit we are considering. Notice that a $(0,1)$-sevenbrane induces the following SL$(2, \zz)$ monodromy on the axio-dilation:
\be
\tau \mapsto \frac{a \tau + b}{c \tau +d}\,, \qquad {\rm with} \quad \begin{pmatrix} a & b\\c & d \end{pmatrix} = \begin{pmatrix} 1 & 0 \\ -1 & 1 \end{pmatrix}\,.
\ee
This will trigger an S-duality that sends $g_s \mapsto \infty$. Hence, we conclude that our weak coupling limit somehow pushes these branes away, keeping NS5-branes non-compact, and hence non-dynamical.

Can we still perform the alignement of the color and flavor branes and lift the NS5-brane as discussed? Yes. However, the interpretation is that we are switching on a Fayet-Iliopoulos term (as was originally suspected in \cite{Aharony:1997bh}).

\subsection{Eliminating $U(1)$ factors by compactifying 5-brane webs}
Having argued how 5-brane webs give rise to unitary quiver gauge theories when the string coupling is zero, we should now be able to mimick the St\"uckelberg mechanism we saw in IIA when we compactified the direction $x^5$ on a circle. In the 5-brane web diagram, this will simply correspond to compactifying the vertical coordinate of the $(p, q)$-plane. This compactification, and the no-bending regime, are summarized in Table~\ref{tab:5branesunusual}.
\begin{table}[h!]
\be
\begin{tabular}{c|cccccccccc}
 & 0 & 1 & 2 & 3 & 4 & 5 & 6 & 7 & 8 & 9\\
D5 & --- & --- & --- & --- & ---  &   &  --- & &  &  \\
 NS5 & --- & --- & --- & --- & ---  &   --- &    &   & &  \\ 
(1,1) 5-branes & --- & --- & --- & --- & ---     &    \multicolumn{2}{c}{angle} & & & \\ 
 7-branes & --- & --- & --- & --- & ---  &   && ---    &  --- & ---\\
 &  &  &  &  &   & $S^1$  && && $S^1$ 
 \end{tabular}\nonumber
 \ee
\caption{Brane setup when one compactifies the $x^8$ direction.} \label{tab:5branesunusual}
\end{table}
How can we see that this will turn $U(N)$ groups into $SU(N)$ groups? We will see this indirectly by inspecting the Higgs branch.

One difference between unitary and special unitary groups is the fact that the former do not have baryonic branches, whereas the latter do. As we explained in the previous section, baryonic branches correspond to aligning $N_c$ flavor branes on one side with all $N_c$ color branes, and then simply lift the separating NS5-brane off the plane, which itself must be suspended between $(0,1)$-sevenbranes in order to make for a normalizable move. However, in our weak coupling limit, such sevenbranes are banned.

On the other hand, having compactified precisely the vertical direction of the $(p,q)$-plane, NS5-branes (all vertical branes) are rendered truly five-dimensional. This means that lifting such branes off the plane can be interpreted as a local move, i.e. giving a vev to a dynamical field (this in fact is dual in IIA to a  complex structure modulus of the local K3, which is dynamical after compactifying $x^5$ on a circle). Indeed, this explains the enhancement of the Higgs branch when going from $U(N)$ to $SU(N)$ gauge groups. This move is now fully compatible with our weakly coupled regime.

\section{Discussion}
In this paper, we studied the Higgs branches of quiver gauge theories obtained by putting M-theory on $\cC$-fibrations over local K3's modeled as $\cc^2/\Gamma$ for $\Gamma \subset SU(2)$. We found that by counting local deformation moduli, the numbers are consistent with unitary gauge groups, as opposed to the expected special unitary groups.

We also proposed a mechanism to render the $U(1)$ factors massive: Compactifying a transverse direction renders bulk SUGRA multiplets dynamical, and BF-couplings induce the St\"uckelberg mechanism. In M-theory, this translates to having elliptic fibrations over the local K3's.

We gave an interpretation on the dual 5-brane web picture. For finite $g_s^{IIA}$ and a T-duality circle $S^1_{IIA}$ of radius going to infinity, $g_s^{IIB} \rightarrow 0$. In this case, brane bending is suppressed, and D5-segments suspended between parallel NS5's are free to move up and down (as in 3d Hanany-Witten constructions), thereby making the enhanced Coulomb branch visible.

Several puzzles remain:
\begin{itemize}
\item The normalizable two-forms in M-theory will become non-normalizable if we take an ALE limit of the fiber ALF metric. In this case, we expect to return to the $SU$-quiver situation, without the need for the perturbative St\"uckelberg mechanism. We do not know of a perturbative IIA explanation for this. Presumably, the D6-branes are becoming delocalized at strong $g_s$, and their worldvolume photons are behaving more and more like bulk fields which are no longer 5d.

\item Algebraic spaces of the form $uv = \Delta$ are compatible with ALF as well as ALE spaces. However, when we count complex structure moduli in algebraic geometry, it seems that we are seeing unitary gauge groups, which are supported only by ALF spaces. This is intriguing, and begs the question, whether non-algebraic moduli might be lurking (from the base manifold), in the ALE limit.

\item We do not know, whether UV fixed-points see a difference between having these Abelian factors or not, as we have not attempted to compute infinite coupling Higgs branches, as is done in \cite{Ferlito:2017xdq,Cremonesi:2015lsa,Cabrera:2018jxt, Bourget:2019aer} among other works. One possibility is that our theories do not have UV fixed points, but are UV completed by little string theories, since the asymptotic radius of the ALF-fibration keeps a scale in the game\footnote{We thank M. Del Zotto for suggesting this possibility to us.}.

\item All of our setups with balanced quiver nodes can be obtained from 6d via KK reduction. However, those are not the so-called `KK theories' discussed in the literature. For instance, in order to obtain $SU(2)$ with eight flavors, one would have to introduce orientifolds.

\end{itemize}

%%%%%%%%%%%%%%%%%%%%%%%%%%%%%%%%%%%%%%%%%%%%%%%%
\section*{Acknowledgments} % sec (acknow)
%%%%%%%%%%%%%%%%%%%%%%%%%%%%%%%%%%%%%%%%%%%%%%%%

We have benefited from discussions with Riccardo Argurio, Antoine Bourget, Cyril Closset, Michele Del Zotto, Lorenzo Di Pietro, Julius Grimminger, Eduardo Garc\'ia-Valdecasas, Vivek Saxena, S.~Schäfer-Nameki, Marcus Sperling, Valdo Tatitscheff. A.C.~is a Research Associate of the Fonds de la Recherche Scientifique F.N.R.S.~(Belgium). The work of A.C.~is partially supported by IISN - Belgium (convention 4.4503.15), and supported by the Fonds de la Recherche Scientifique - F.N.R.S.~under Grant CDR J.0181.18. 
The work of R.V.~is partially supported by ``Fondo per la Ricerca di Ateneo - FRA 2018'' (UniTS) and by INFN Iniziativa Specifica ST\&FI.

% fold sec (acknow)

%%%%%%%%%%%%%%%%%%%%%%%%%%%%%%%%%%%%%%%%%%%%%%%%

%%%%%%%%%%%%%%%%%%%%%%%%%%
% BIBLIOGRAPHY
%%%%%%%%%%%%%%%%%%%%%%%%%%
\providecommand{\href}[2]{#2}

\bibliographystyle{at}

\begin{thebibliography}{10}

\bibitem{Intriligator:1997pq}
K.~A. Intriligator, D.~R. Morrison, and N.~Seiberg, ``{Five-dimensional
  supersymmetric gauge theories and degenerations of Calabi-Yau spaces},'' {\em
  Nucl. Phys.} {\bf B497} (1997) 56--100,
\href{http://arXiv.org/abs/hep-th/9702198}{{\tt hep-th/9702198}}.
%%CITATION = HEP-TH/9702198;%%.

\bibitem{Seiberg:1996bd}
N.~Seiberg, ``{Five-dimensional SUSY field theories, nontrivial fixed points
  and string dynamics},'' {\em Phys. Lett.} {\bf B388} (1996) 753--760,
\href{http://arXiv.org/abs/hep-th/9608111}{{\tt hep-th/9608111}}.
%%CITATION = HEP-TH/9608111;%%.

\bibitem{Apruzzi:2019kgb}
F.~Apruzzi, S.~Schafer-Nameki, and Y.-N. Wang, ``{5d SCFTs from Decoupling and
  Gluing},'' \href{http://arXiv.org/abs/1912.04264}{{\tt 1912.04264}}.

\bibitem{Apruzzi:2019vpe}
F.~Apruzzi, C.~Lawrie, L.~Lin, S.~Sch{\"a}fer-Nameki, and Y.-N. Wang, ``{5d
  Superconformal Field Theories and Graphs},'' {\em Phys. Lett. B} {\bf 800}
  (2020) 135077, \href{http://arXiv.org/abs/1906.11820}{{\tt 1906.11820}}.

\bibitem{Bhardwaj:2020gyu}
L.~Bhardwaj and G.~Zafrir, ``{Classification of 5d N=1 gauge theories},''
  \href{http://arXiv.org/abs/2003.04333}{{\tt 2003.04333}}.

\bibitem{Bhardwaj:2018vuu}
L.~Bhardwaj and P.~Jefferson, ``{Classifying 5d SCFTs via 6d SCFTs: Arbitrary
  rank},'' {\em JHEP} {\bf 10} (2019) 282,
  \href{http://arXiv.org/abs/1811.10616}{{\tt 1811.10616}}.

\bibitem{Bhardwaj:2018yhy}
L.~Bhardwaj and P.~Jefferson, ``{Classifying 5d SCFTs via 6d SCFTs: Rank
  one},'' {\em JHEP} {\bf 07} (2019) 178,
  \href{http://arXiv.org/abs/1809.01650}{{\tt 1809.01650}}. [Addendum: JHEP 01,
  153 (2020)].

\bibitem{Bhardwaj:2019xeg}
L.~Bhardwaj, ``{Do all $5d$ SCFTs descend from $6d$ SCFTs?},''
  \href{http://arXiv.org/abs/1912.00025}{{\tt 1912.00025}}.

\bibitem{Apruzzi:2019opn}
F.~Apruzzi, C.~Lawrie, L.~Lin, S.~Sch{\"a}fer-Nameki, and Y.-N. Wang, ``{Fibers
  add Flavor, Part I: Classification of 5d SCFTs, Flavor Symmetries and BPS
  States},'' {\em JHEP} {\bf 11} (2019) 068,
  \href{http://arXiv.org/abs/1907.05404}{{\tt 1907.05404}}.

\bibitem{Apruzzi:2019enx}
F.~Apruzzi, C.~Lawrie, L.~Lin, S.~Sch{\"a}fer-Nameki, and Y.-N. Wang, ``{Fibers
  add Flavor, Part II: 5d SCFTs, Gauge Theories, and Dualities},'' {\em JHEP}
  {\bf 03} (2020) 052, \href{http://arXiv.org/abs/1909.09128}{{\tt
  1909.09128}}.

\bibitem{Closset:2018bjz}
C.~Closset, M.~Del~Zotto, and V.~Saxena, ``{Five-dimensional SCFTs and gauge
  theory phases: an M-theory/type IIA perspective},'' {\em SciPost Phys.} {\bf
  6} (2019), no.~5, 052,
\href{http://arXiv.org/abs/1812.10451}{{\tt 1812.10451}}.
%%CITATION = ARXIV:1812.10451;%%.

\bibitem{Morrison:2020ool}
D.~R. Morrison, S.~Schafer-Nameki, and B.~Willett, ``{Higher-Form Symmetries in
  5d},'' \href{http://arXiv.org/abs/2005.12296}{{\tt 2005.12296}}.


\bibitem{Albertini:2020mdx}
F.~Albertini, M.~Del Zotto, I.~Garcia Etxebarria and S.~S.~Hosseini,
``{Higher Form Symmetries and M-theory},''
\href{http://arXiv.org/abs/2005.12831}{{\tt 2005.12831}}.

\bibitem{Jefferson:2018irk}
P.~Jefferson, S.~Katz, H.-C. Kim, and C.~Vafa, ``{On Geometric Classification
  of 5d SCFTs},'' {\em JHEP} {\bf 04} (2018) 103,
  \href{http://arXiv.org/abs/1801.04036}{{\tt 1801.04036}}.

\bibitem{Bhardwaj:2019jtr}
L.~Bhardwaj, ``{On the classification of $5d$ SCFTs},''
  \href{http://arXiv.org/abs/1909.09635}{{\tt 1909.09635}}.

\bibitem{Saxena:2019wuy}
V.~Saxena, ``{Rank-two 5d SCFTs from M-theory at isolated toric singularities:
  a systematic study},''
\href{http://arXiv.org/abs/1911.09574}{{\tt 1911.09574}}.
%%CITATION = ARXIV:1911.09574;%%.

\bibitem{Jefferson:2017ahm}
P.~Jefferson, H.-C. Kim, C.~Vafa, and G.~Zafrir, ``{Towards Classification of
  5d SCFTs: Single Gauge Node},'' \href{http://arXiv.org/abs/1705.05836}{{\tt
  1705.05836}}.

\bibitem{Eckhard:2020jyr}
J.~Eckhard, S.~Schafer-Nameki, and Y.-N. Wang, ``{Trifectas for $T_N$ in 5d},''
  \href{http://arXiv.org/abs/2004.15007}{{\tt 2004.15007}}.

\bibitem{Closset:2019juk}
C.~Closset and M.~Del~Zotto, ``{On 5d SCFTs and their BPS quivers. Part I:
  B-branes and brane tilings},'' \href{http://arXiv.org/abs/1912.13502}{{\tt
  1912.13502}}.

\bibitem{Apruzzi:2018nre}
F.~ Apruzzi, L.~ Lin, and C.~ Mayrhofer, ``{Phases of 5d SCFTs from M-/F-theory on Non-Flat Fibrations},'' {\em JHEP} {\bf 05} (2019) 187,
  \href{http://arXiv.org/abs/1811.12400}{{\tt 1811.12400}}.
  
\bibitem{Ruback:1986ag}
P.~Ruback, ``{The Motion of {Kaluza-Klein} Monopoles},'' {\em Commun. Math.
  Phys.} {\bf 107} (1986) 93--102.

\bibitem{Hanany:1997gh}
A.~Hanany and A.~Zaffaroni, ``{Branes and six-dimensional supersymmetric
  theories},'' {\em Nucl. Phys. B} {\bf 529} (1998) 180--206,
  \href{http://arXiv.org/abs/hep-th/9712145}{{\tt hep-th/9712145}}.

\bibitem{Brunner:1997gf}
I.~Brunner and A.~Karch, ``{Branes at orbifolds versus Hanany Witten in
  six-dimensions},'' {\em JHEP} {\bf 03} (1998) 003,
  \href{http://arXiv.org/abs/hep-th/9712143}{{\tt hep-th/9712143}}.

\bibitem{Aganagic_2010}
M.~Aganagic, ``A stringy origin of m2 brane chern--simons theories,'' {\em
  Nuclear Physics B} {\bf 835} (Aug, 2010) 1--28.

\bibitem{Benini:2009qs}
F.~Benini, C.~Closset, and S.~Cremonesi, ``{Chiral flavors and M2-branes at
  toric CY4 singularities},'' {\em JHEP} {\bf 02} (2010) 036,
\href{http://arXiv.org/abs/0911.4127}{{\tt 0911.4127}}.
%%CITATION = ARXIV:0911.4127;%%.

\bibitem{Grimm:2011tb}
T.~W. Grimm, M.~Kerstan, E.~Palti, and T.~Weigand, ``{Massive Abelian Gauge
  Symmetries and Fluxes in F-theory},'' {\em JHEP} {\bf 12} (2011) 004,
\href{http://arXiv.org/abs/1107.3842}{{\tt 1107.3842}}.
%%CITATION = ARXIV:1107.3842;%%.

\bibitem{Etesi:2008ew}
G.~Etesi and S.~Szabo, ``{Harmonic functions and instanton moduli spaces on the
  multi-Taub-NUT space},'' {\em Commun. Math. Phys.} {\bf 301} (2011) 175--214,
  \href{http://arXiv.org/abs/0809.0480}{{\tt 0809.0480}}.

\bibitem{Braun_2014}
A.~P. Braun, A.~Collinucci, and R.~Valandro, ``The fate of u(1)'s at strong
  coupling in f-theory,'' {\em Journal of High Energy Physics} {\bf 2014} (Jul,
  2014).

\bibitem{Hanany:1996ie}
A.~Hanany and E.~Witten, ``{Type IIB superstrings, BPS monopoles, and
  three-dimensional gauge dynamics},'' {\em Nucl. Phys. B} {\bf 492} (1997)
  152--190, \href{http://arXiv.org/abs/hep-th/9611230}{{\tt hep-th/9611230}}.

\bibitem{Aharony:1997bh}
O.~Aharony, A.~Hanany, and B.~Kol, ``{Webs of (p,q) five-branes,
  five-dimensional field theories and grid diagrams},'' {\em JHEP} {\bf 01}
  (1998) 002, \href{http://arXiv.org/abs/hep-th/9710116}{{\tt hep-th/9710116}}.

\bibitem{Ferlito:2017xdq}
G.~Ferlito, A.~Hanany, N.~Mekareeya, and G.~Zafrir, ``{3d Coulomb branch and 5d
  Higgs branch at infinite coupling},'' {\em JHEP} {\bf 07} (2018) 061,
  \href{http://arXiv.org/abs/1712.06604}{{\tt 1712.06604}}.

\bibitem{Cremonesi:2015lsa}
S.~Cremonesi, G.~Ferlito, A.~Hanany, and N.~Mekareeya, ``{Instanton Operators
  and the Higgs Branch at Infinite Coupling},'' {\em JHEP} {\bf 04} (2017) 042,
  \href{http://arXiv.org/abs/1505.06302}{{\tt 1505.06302}}.

\bibitem{Cabrera:2018jxt}
S.~Cabrera, A.~Hanany, and F.~Yagi, ``{Tropical Geometry and Five Dimensional
  Higgs Branches at Infinite Coupling},'' {\em JHEP} {\bf 01} (2019) 068,
  \href{http://arXiv.org/abs/1810.01379}{{\tt 1810.01379}}.

\bibitem{Bourget:2019aer}
A.~Bourget, S.~Cabrera, J.~F. Grimminger, A.~Hanany, M.~Sperling, A.~Zajac, and
  Z.~Zhong, ``{The Higgs mechanism --- Hasse diagrams for symplectic
  singularities},'' {\em JHEP} {\bf 01} (2020) 157,
  \href{http://arXiv.org/abs/1908.04245}{{\tt 1908.04245}}.

\end{thebibliography}

\end{document}